\documentclass[10pt,prl,twocolumn,notitlepage,floatfix,superscriptaddress]{revtex4}
\usepackage{graphicx}
\usepackage{bbm}
\usepackage{epstopdf}
\usepackage[colorlinks]{hyperref}
\hypersetup{
 colorlinks=true,
 citecolor=black,
 linkcolor=black,
 urlcolor=black}
\usepackage{amssymb}
\usepackage{times}
\usepackage{amsmath,mathrsfs}
\usepackage{siunitx}
\makeatletter
\def\Let@{\def\\{\notag\math@cr}}
\makeatother
\usepackage{amsthm}
\usepackage{amsfonts}%
\usepackage[caption=false]{subfig}
\renewcommand{\sec}[1]{\hyperref[sec:#1]{Sec.~\ref{sec:#1}}}
\newcommand{\etal}{\mbox{\emph{et al.\ }}}
\newcommand{\eq}[1]{(\ref{eq:#1})}
\newcommand{\fig}[1]{\hyperref[fig:#1]{Fig.~\ref{fig:#1}}}
\newcommand{\tab}[1]{\hyperref[fig:#1]{Table.~\ref{tab:#1}}}

\input{Qcircuit}
\graphicspath{{./Figures/}}

\begin{document}
\title{Practical quantum computing on encrypted data}
\author{Kevin Marshall}\thanks{These two authors contributed equally to this work}
\affiliation{Department of Physics, University of Toronto, Toronto, M5S 1A7, Canada}
\author{Christian S. Jacobsen}\thanks{These two authors contributed equally to this work}
\affiliation{Department of Physics, Technical University of Denmark, Fysikvej, 2800 Kgs. Lyngby, Denmark}
\author{Clemens Sch\"afermeier}
\affiliation{Department of Physics, Technical University of Denmark, Fysikvej, 2800 Kgs. Lyngby, Denmark}
\author{Tobias Gehring}
\affiliation{Department of Physics, Technical University of Denmark, Fysikvej, 2800 Kgs. Lyngby, Denmark}
\author{Christian Weedbrook}
\email{christian.weedbrook@gmail.com}
\affiliation{CipherQ, 10 Dundas St.~E, Toronto, M5B 2G9, Canada}
\author{Ulrik L. Andersen}
\affiliation{Department of Physics, Technical University of Denmark, Fysikvej, 2800 Kgs. Lyngby, Denmark}
\begin{abstract}
The ability to perform computations on encrypted data is a powerful tool for protecting a client's privacy, especially in today's era of cloud and distributed computing. In terms of privacy, the best solutions that classical techniques can achieve are unfortunately not unconditionally secure in the sense that they are dependent on a hacker's computational power. Here we theoretically investigate, and experimentally demonstrate with Gaussian displacement and squeezing operations, a quantum solution that achieves the unconditional security of a user's privacy using the practical technology of continuous variables. We demonstrate losses of up to 10 km both ways between the client and the server and show that security can still be achieved. Our approach offers a number of practical benefits, which can ultimately allow for the potential widespread adoption of this quantum technology in future cloud-based computing networks.
\end{abstract}
\date{\today}

\maketitle

\section{Introduction}

Incredibly, 2.5 quintillion bytes of data are produced in the world every single day. In fact, it has been estimated that over 90\% of the world's data was created in the last two years \cite{IBM}. Most of this data is stored around the world in data centers and accessed remotely via the cloud.  Because cloud computing is operated by third parties (e.g., Amazon, Facebook), one of the outcomes of this acceleration in information is the need to better protect our privacy. In principle, the cloud contains various types of data for which security and privacy are essential. For example, an individual's personal data (such as medical records and credit card information), the trade secrets and intellectual property of multinational corporations, and sensitive government information (e.g., the CIA bought cloud space from Amazon). Therefore securing a client's privacy in the cloud is one of the core security challenges we face today. 

One of the current solutions to this challenge is homomorphic encryption \cite{Naehrig11}. The requirement for such a solution was first identified in the 1970's by Rivest and colleagues \cite{Rivest78}. It was not until over 30 years later that IBM researcher Craig Gentry discovered fully homomorphic encryption \cite{Gentry09}. Although there has been much progress in recent years, the best known implementations of fully homomorphic encryption are impractical for today's computers \cite{Naehrig11,Brakerski11,Brakerski12,Gentry15}.

It is interesting when one considers the generalization of homomorphic encryption to the quantum realm.  If one restricts the class of quantum operations to be implemented, then it was shown one can hide up to a constant fraction, which can be made arbitrarily close to unity, of the encrypted information while only requiring polynomial overhead \cite{Tan14}.  Unfortunately, it has been shown that, perfectly secure, deterministic fully homomorphic quantum computation is only possible at the expense of an exponential overhead \cite{Yu14}.  One can relax the requirements of quantum homomorphic encryption by allowing further rounds of interaction between the client and server.  Such a scheme was first studied by Childs~\cite{Childs05} where he outlined a protocol to allow an individual of limited quantum ability (a client) to delegate a quantum computation to another person (a server) who is in possession of a fully-fledged quantum computer. The idea here being that such a computer is only initially available to a select few. Progress towards this direction was demonstrated recently by IBM who made available a small (5 qubit) quantum computer for access in the cloud \cite{IBM2}.  The crux of the issue in developing such a protocol lies in the fact that the client wants to hide some subset of: her input, the quantum program, and the final result.  That such a scheme is even possible is astounding from a classical viewpoint.  Seminal work by Broadbent \etal \cite{Broadbent09} built upon this idea, in the cluster-state framework \cite{Briegel09}, to develop the notion of universal blind quantum computing; a protocol which fulfils all three of the above criteria and requires only that the client is able to prepare and send single qubits from a finite set.

In this paper, we offer a new and novel approach to quantum computing on encrypted data, that does not require  challenging single photon sources or single photon detectors and is based on a different type of substrate, known as continuous variables \cite{Braunstein05,Weedbrook12}. Continuous variables (CVs) offer a number of practical advantages over its qubit counterpart: deterministic gate implementation, low-cost and affordability of components (such as laser sources and detectors), high detection efficiencies, high-rate of information transfer, and the ability to be fully integrated within current telecommunication infrastructures. Here, we answer the questions of how limited the client's quantum capacity can be, how much classical communication is required between the client and server, and how many classical/quantum operations are needed per gate.  Furthermore, we provide proof-of-principle experimental results which highlight the effect of loss over 10~km of effective loss.

\begin{figure}[hbtp]
\centering
\includegraphics[width=8.5cm]{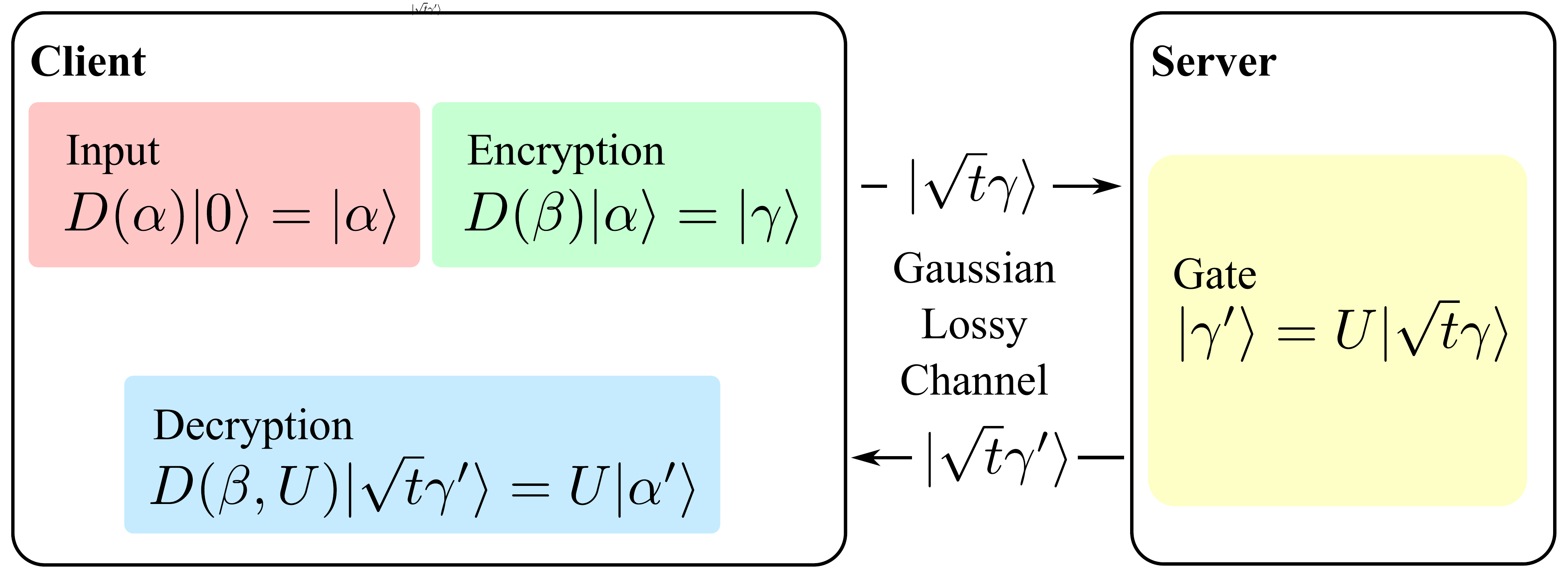}
\caption{\textbf{Protocol for quantum computing on encrypted data.} Input: A displaced vacuum state is prepared. Encryption: A random displacement is applied to the initial state as an encryption procedure. Channel: The state is transmitted over a Gaussian lossy channel to the server (transmission $t$). Gate: The server applies the desired (Gaussian displacement or squeezing) unitary. Channel: The state is sent back over the Gaussian lossy channel to the client. Decryption: The client applies a decryption operation to retrieve the final output state.}
\label{fig:Scheme}
\end{figure}
Our protocol for computing on encrypted CVs consists of three stages (cf.\ \fig{Scheme}): an encryption stage, a program stage and a decryption stage.  We will now elaborate in more detail. First, the client performs an encryption operation on their desired input to limit the amount of information the server can obtain about the initial state.  The state is then sent to the server, who performs a predetermined set of gates known to both parties (corresponding to the program needed to be performed).  Finally, the state is sent back to the client who is able to perform a decryption operation which recovers the output of the desired computation.  

To discuss the encryption operation we first define the Heisenberg-Weyl operators \cite{Weedbrook12} $X(Q)=\exp(-iQ\hat p)$ and $Z(P)=\exp(iP\hat q)$, as well as the displacement operator $D(\alpha)=\exp(\alpha \hat a^\dagger - \alpha^* \hat a)$ where $\hat q,\hat p$ are the canonical amplitude and phase operators, respectively, which obey Heisenberg's uncertainty relation $[\hat q,\hat p]=i$.  The annihilation and creation operators are denoted by $\hat a,\hat a^\dagger$ respectively, and are defined by $\hat a=(\hat q+i\hat p)/\sqrt 2$ and its adjoint.  Consider the action of applying a random displacement in phase-space to the input state; intuitively speaking, if the displacement is chosen randomly then the state will look mixed and smeared out over all of phase-space to somebody unaware of the displacement parameters.  This can be made rigorous by invoking the fact that $\frac{1}{\pi}\int D(\alpha)\ket\psi\bra\psi D^{\dagger}(\alpha)d^2\alpha =\mathbbm{1}$
holds for any normalized state $\ket\psi$.  This shows us that averaging the displacements applied to any normalized state, over the entire complex plane, results in a quantity proportional to the identity.  In reality, a uniform distribution of displacements over $\mathbb R^2$ is unphysical and we must replace this with a function which dies off sufficiently fast in order to adhere to some energy threshold.  For a fixed energy, a Gaussian distribution will maximize the entropy of the resulting state \cite{Holevo99} and thus we restrict ourselves to Gaussian distributions.  This will potentially give the server the capability of extracting some information about the input state, but the amount will be bounded based on the width of the Gaussian.  A formal security analysis of our protocol, in the limit of finite squeezing and displacements, remains an open problem.

We now turn our attention to the server who is asked to perform some known algorithm on the encrypted data, where in principle, the algorithm corresponds to a universal quantum computation. To show that the server can perform such a computation, it suffices to show that we can implement a universal set of gates. Namely, we need to show that the set $\mathcal G=\{X(Q),Z(P),U_2(T),U_3(T),F,C_Z\}$ where $U_k(T)=\exp(iT\hat q^k)$, $F=\exp\left[\frac{i\pi}{4}\left(\hat q^2+\hat p^2\right)\right]$, and $C_Z=\exp(i\hat q_1\otimes\hat q_2)$ can be implemented and decrypted with an appropriately chosen operation. Note that of the gates in this set only the $U_3(T)$ gate \cite{Marek11,Marshall15} is not Gaussian, and we only require this one non-Gaussian gate in order to achieve universal quantum computation \cite{Menicucci06}.  However, it is prudent to note that non-Gaussian operations are very challenging to implement and that there remain many challenges in devising a robust quantum computer based on continuous variables \cite{Menicucci06}.
\begin{table}[htp]
\begin{tabular}{l|l}
Gate & Correction \\ 
\hline 
$Z(T)$ & $X(-Q)Z(-P)$\\
$X(T)$ & $X(-Q)Z(-P)$\\
$U_2(T)$ & $X(-Q)Z(-2QT-P)$\\
$U_3(T)$ & $X(-Q)Z(3Q^2T-P)U_2(-3QT)$\\
$F$ & $X(P)Z(-Q)$\\
$C_Z$ & $X_1(-Q_1)Z_1(-Q_2-P_1)\otimes X_2(-Q_2)Z_2(-Q_1-P_2)$
\end{tabular}
\caption{The decryption operations corresponding to each gate, up to a phase, for the encryption operation $D(Q,P)$ for single mode gates and $D_1(Q_1,P_1)D_2(Q_2,P_2)$ for two-mode gates.}
\label{tab:unlocking}
\end{table}
Except for $U_3$, all of these operations have decryption operators that correspond to displacements (cf. Table~\ref{tab:unlocking}), and this allows for the straightforward composition of gates. To make it clearer we consider a simple example, namely the $Z(S)$ gate.  The encryption operation $D(Q,P)$ consists of a translation in both the amplitude and phase quadratures and it can be decomposed as a sequence of an $X(Q)$ as well as a $Z(P)$ gate, the latter commutes with $Z(S)$ and so will simply slide through the gate.  Consider the application of the $X(Q)$ gate; this gate slides through the $Z(S)$ gate up to a phase as $X(Q)Z(S)=e^{-iQS}Z(S)X(Q)$.
Thus we can construct a decryption operation as
\begin{align}
Z(S)D(Q,P)&=C^\dagger(Q,P,S)Z(S),
\end{align}
where $C(Q,P,S)=\exp[i(QP/2-QS)]X(-Q)Z(-P)$ is the decryption gate which, when applied, will undo the effect of the initial encryption operation.  However, the $U_2$ decryption gate present in the $U_3$ operation does not easily slide through the Fourier gate $F$, and thus we must have the server correct for this on-the-fly; this is possible in a manner similar to the discrete-variable protocol presented in Ref.~\cite{Fisher14}.  To perform $U_3(T)$ the client instead sends the server two modes, the first of which is the encrypted state and the second being the state $U_2(A)Z(Q')\ket 0_p$, where $A, Q'$ are chosen randomly and $\ket 0_p$ denotes a momentum eigenstate; finite squeezing does not present any issues other than the ones normally associated with teleportation, namely the introduction of extra Gaussian noise \cite{Braunstein98}. The server is then able to implement $U_3(T)$ as shown in the circuit below.
\begin{align*}
\Qcircuit @C=1em @R=1em {
\push{\rule{6em}{0em}} & \lstick{D(Q,P)\ket\psi} &\gate{F^\dagger U_3(T)}& \ctrl{1} & \qw &\measureD{\hat p=m_1}\\
\push{\rule{6em}{0em}} & \lstick{U_2(A)Z(Q')\ket{0}_p} &\qw& \ctrl{-1} & \gate{D(Q'',P'')}& \gate{U_2(B)},
}
\end{align*}
After the client sends both modes, and the value of $B$, the server performs the desired $U_3(T)$ gate, after application of the inverse Fourier gate, before interacting the two modes with a controlled phase gate, indicated by the vertical line.  The server then measures the first mode and after performing the $U_2(B)$ gate obtains the desired state $U_3(T)\ket\psi$ on the remaining mode, up to displacement corrections, provided that $A+B=-3QT$.  Note that these additional corrections depend on the value $m_1A$ and so the server must communicate the value of $m_1$ obtained to the client, thus requiring one round of classical communication for the full implementation.


We discuss the implementation of this protocol in more detail in the supplementary material, including a discussion of the decryption operations for each gate in the universal set, how to compose gates, the effects of transmission and imperfect encryption, an entanglement-based analogue, the use of channel estimation and limitations on squeezing.

\section{Experiment}

The CV quantum gates associated with linear phase space displacements and squeezing transformations allow for an experimental test of quantum computing on encrypted data solely based on Gaussian quantum states and Gaussian operations.
Such operations can be performed with high fidelity within the field of CV quantum optics.
In the following, we thus use Gaussian displacement and squeezing operations to test the basic principles of quantum computing on encrypted data, i.e.\ we implement a server performing firstly $Z$ and $X$ gates and secondly a squeezing gate, related to the $U_2$ gate.

\begin{figure}[tbhp]
\centering
\subfloat{\includegraphics[width=8.5cm]{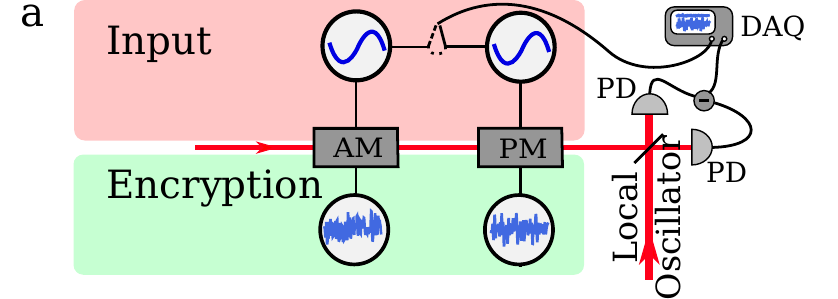}\label{fig:MutualSetup}}\\
\subfloat{\includegraphics[width=8.5cm]{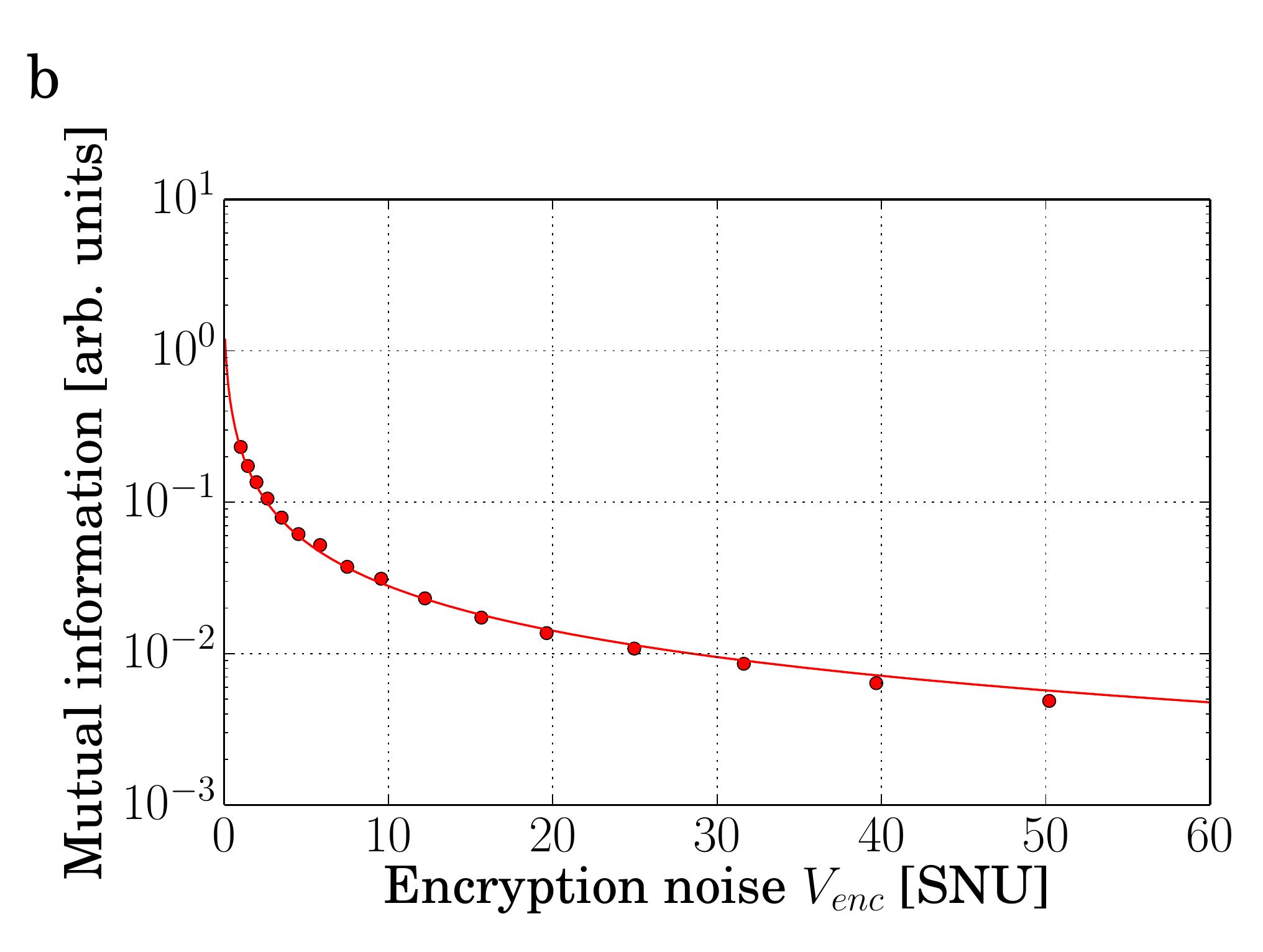}\label{fig:Mutual}}
\caption{{\bf Effectiveness of encryption}. (a) Experimental setup using electro-optical amplitude (AM) and phase (PM) modulators driven by Gaussian white noise sources to generate ensembles of encrypted coherent states. Both the input noise and the homodyne output were demodulated at 10.5\,MHz and recorded by a data acquisition system (DAQ). PD: Photo Detectors. (b) The mutual information $I(\text{server}_\text{enc}\!:\!\text{client}_\text{in})$ for a coherent state chosen according to a Gaussian alphabet with variance $V_\text{in}=0.6$\, shot noise units (SNU), which is then encrypted with a varied encryption variance $V_\text{enc}$. For a fixed distribution of input states the plot shows how an increased encryption noise decreases the server's knowledge of the inputs. Error bars are smaller than the point size.}
\label{fig:EncryptionEffectiveness}
\end{figure}

We start by testing the effectiveness of the encryption operation by using the experimental setup shown in Fig.~\ref{fig:MutualSetup}.
Quantum information at the location of the client was generated in the form of a coherent state of light $\ket\phi = \ket\alpha$.
To test the protocol for many different coherent state excitations simultaneously, we produced an ensemble of coherent states by means of a set of electro-optical modulators (EOMs), thereby preparing the Gaussian ensemble $\rho=\int G_\text{in}(Q,P)D(Q,P)\ket 0\bra 0 D^\dagger (Q,P) dQdP$ where $G_\text{in}(Q,P)$ is a Gaussian probability density function with variance $V_\text{in}$.
This information was then encrypted by applying a randomized phase space displacement onto the coherent state ensemble using the same two EOMs driven by two independent Gaussian white noise sources with equal variances $V_Q=V_P=V_\text{enc}$ for the amplitude ($Q$) and phase ($P$) quadratures.
This encryption noise results in an encrypted state $\rho=\int G_\text{tot}(Q,P) D(Q,P)\ket 0\bra 0 D(Q,P)^\dagger dQdP$ where $G_\text{tot}(Q,P)$ is a Gaussian distribution with a total variance of $V_\text{in}+V_\text{enc}$.
We measured the encrypted quantum states with homodyne detection, and recorded the correlations between the measurements and the input signal. Using these correlations we calculate the mutual information as plotted in Fig.~\ref{fig:Mutual}. The solid line is a theoretical prediction given by
\begin{equation}
I(\text{server}_\text{enc}\!:\!\text{client}_\text{in}) = \dfrac{1}{2} \ln \left(1+ \dfrac{V_\text{in}}{V_\text{enc}} \right)\ .
\label{eq:mutualinformation}
\end{equation}

For efficient encryption, the quadrature correlations and, thus, the mutual information between the encrypted state received by the server and the input state prepared by the client should be vanishingly small.
From the plot, we clearly see the effects of a finite encryption variance.

\begin{figure*}
\centering
\subfloat{\includegraphics[width=4.5cm]{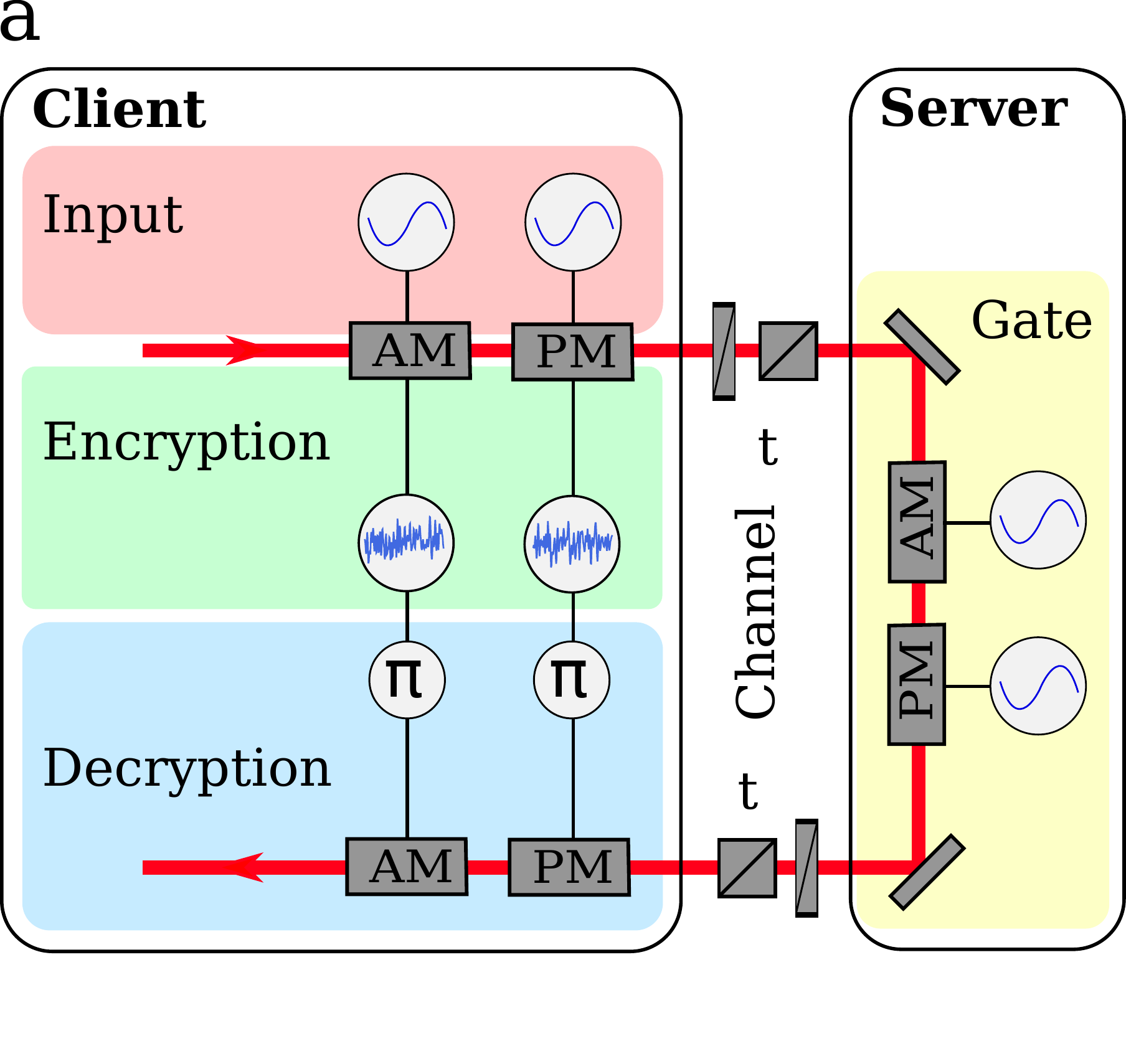}\label{fig:SetupDisplacements}}
\subfloat{\includegraphics[width=6.3cm]{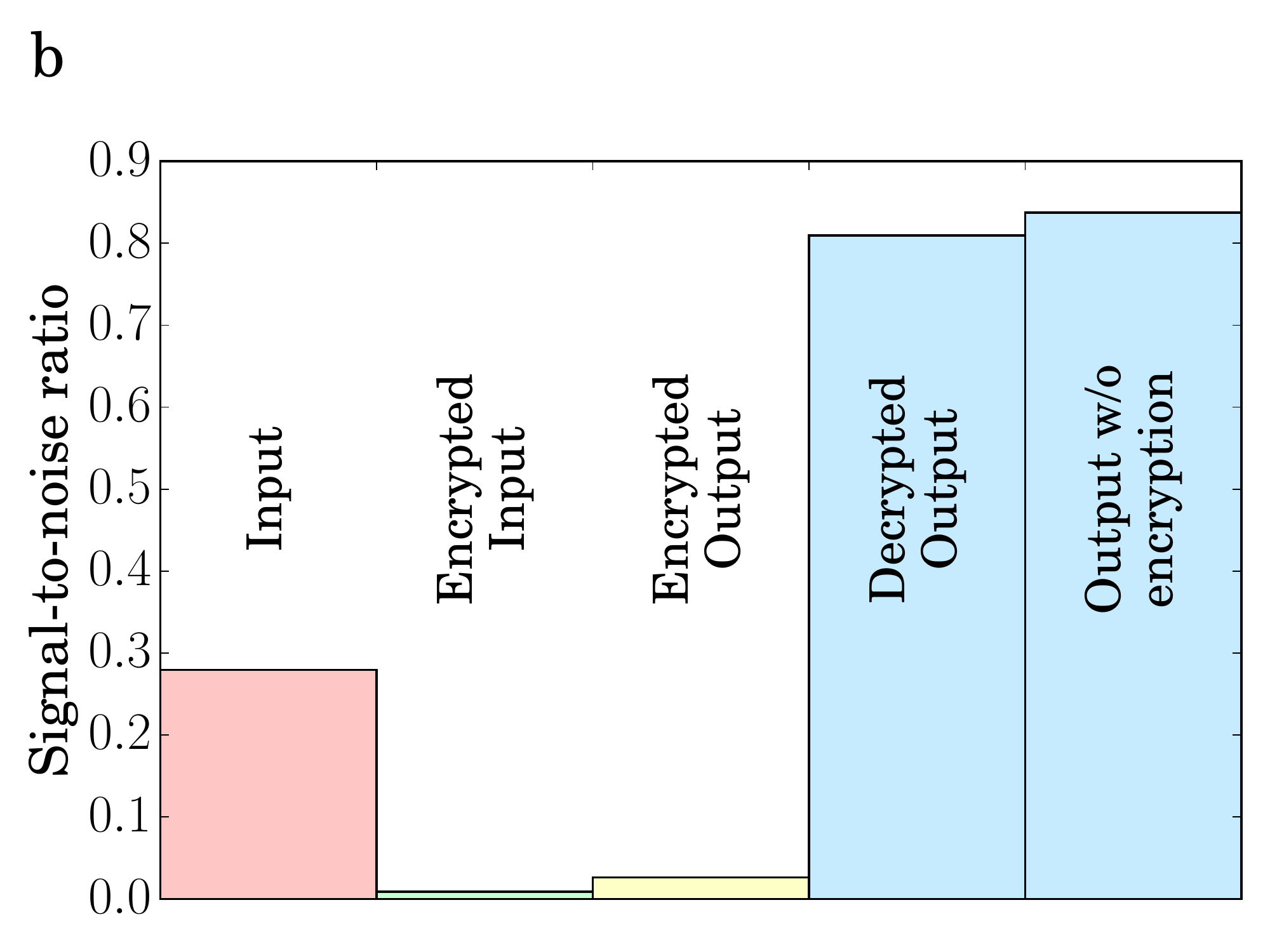}\label{fig:CoherentSNR}}
\hspace*{0cm}
\subfloat{\includegraphics[width=6.3cm]{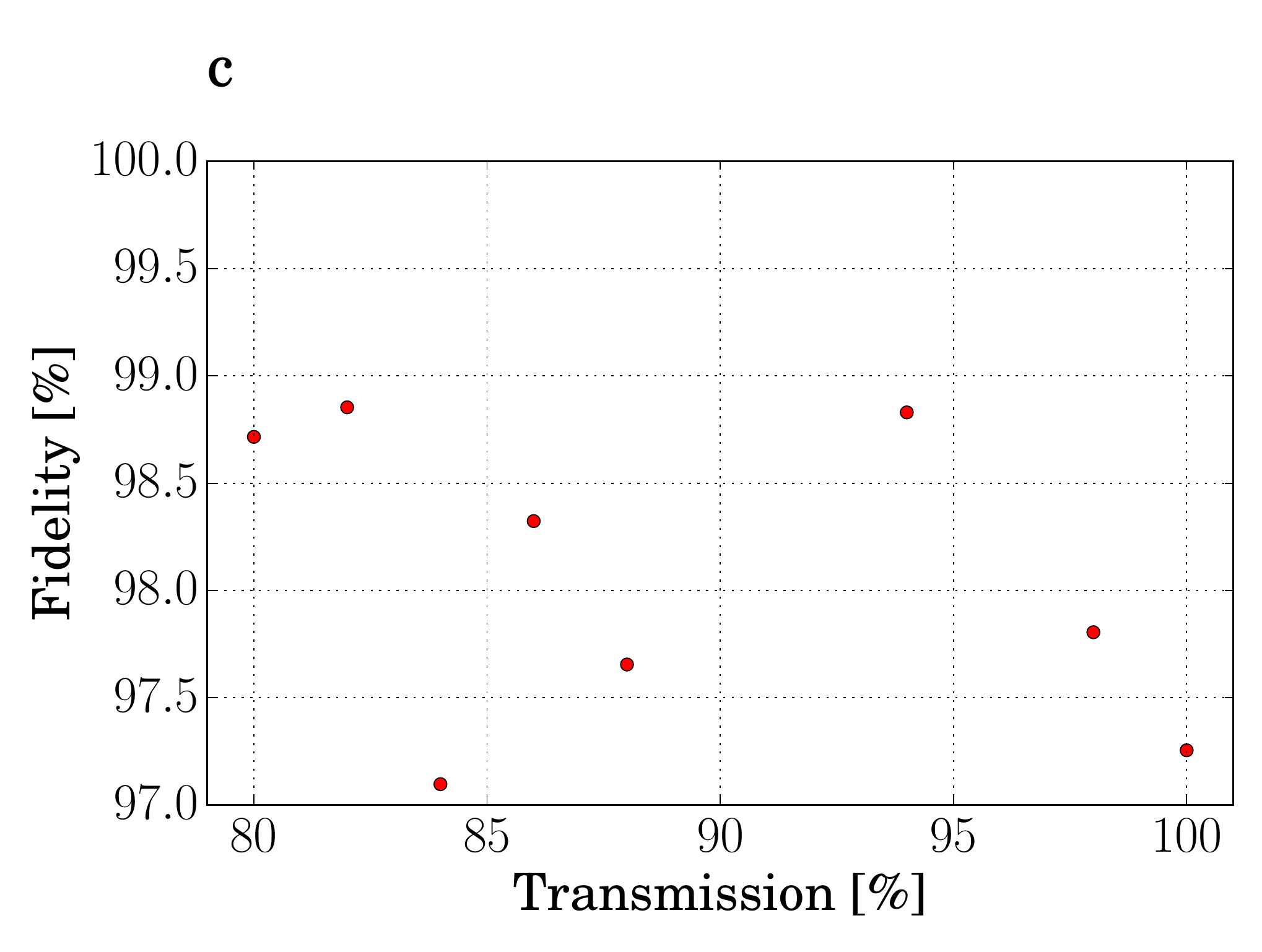}\label{fig:FidelitiesDisplacements}}
\caption{{\bf Server implementing $Z$ and $X$ gates.} (a) Experimental setup. Ensembles of encrypted input states were generated by two electro-optical modulators driven by two Gaussian white noise sources. The encrypted quantum states were then sent to the server through a channel with transmittivity $T$ simulated by a half-wave plate and a polarizing beam splitter. The server performed ensembles of displacements by using a second pair of modulators driven by independent Gaussian white noise generators. After sending the quantum state back to the client the output of the quantum computer was decrypted by applying phase shifted modulations using the encryption noise only known to the client.  (b) The signal-to-noise ratios of the phase quadrature measured by homodyne detection behind each stage of the protocol. An ensemble of coherent input states and gate displacements were used and the transmissivity of the channels was set to $1$. The output of the homodyne detector was recorded by a spectrum analyzer measuring zero-span around $10.5$\,MHz with a resolution bandwidth of $300$\,kHz and a video bandwidth of $30$\,Hz. A small input state was prepared (red) and subsequently encrypted (green). Then a small displacement was performed by the server (yellow), acting as the gate. Afterwards the state was returned to the client for decryption (blue) which yields the output of the quantum computation. For comparison we have recorded the outcome without encrypting the input (upper trace). The difference between these is indicative of the loss for signal-to-noise ratio from imperfect decryption. (c) Fidelities between the output state ensembles using quantum computation on encrypted states and using quantum computation on plain-text states versus the channel transmission $T$. Statistical error bars smaller than the point size. The variation in the fidelities comes mainly from systematic errors in the fine tuning of the phase and gain settings of the decryption noise for optimal decryption.}
\label{fig:displacementgate}
\end{figure*}


We first implement the $Z$ and $X$ displacement gates as illustrated in Fig.~\ref{fig:SetupDisplacements}. The protocol was performed with a Gaussian alphabet of coherent states with variance $V_\text{in}=0.28$ shot-noise units (SNU) embedded in encryption noise of $V_\text{enc}=31$ SNU. For this particular encryption the mutual information is $I=\SI{0.005}{bits/use}$. The $Z$, $X$ gates were tested for a symmetric Gaussian distribution of displacements with variance $V_\text{gate} = 0.6$ SNU. This results in the state $\rho$ after the computation having a total variance of $V_\text{gate}+V_\text{in}+V_\text{enc}$.

Finally, the state is sent back through a lossy channel to the client who is decrypting the state using two EOMs driven by noise which was optimally anti-correlated with the noise used for encryption.
The final state is then ideally given by $D(Q,P) \ket\phi $, but due to imperfections there is some residual noise from the encryption protocol and thus we must consider the output state $\rho$ with a variance of $V_\text{in}+V_\text{gate}+V_\text{res}$.
For the presented measurement the residual noise was $V_\text{res} = 0.072$ SNU.
To visualize the evolution of the information content at different stages of the scheme, in Fig.~\ref{fig:CoherentSNR}, we plot the signal-to-noise-ratios (SNRs) of a single quadrature after each stage of the protocol. It is clear from these numbers that the amount of information in the encrypted state is close to zero, and that the decryption operation is almost ideal.  
For further quantification we show in Fig.~\ref{fig:FidelitiesDisplacements} the fidelity between the ensembles of output states of the quantum computation using encrypted states and plain-text states.
The fidelities are above $97\,\%$ for all measured transmission values. 

\begin{figure*}
    \centering
    \subfloat{\includegraphics[width=0.4\textwidth]{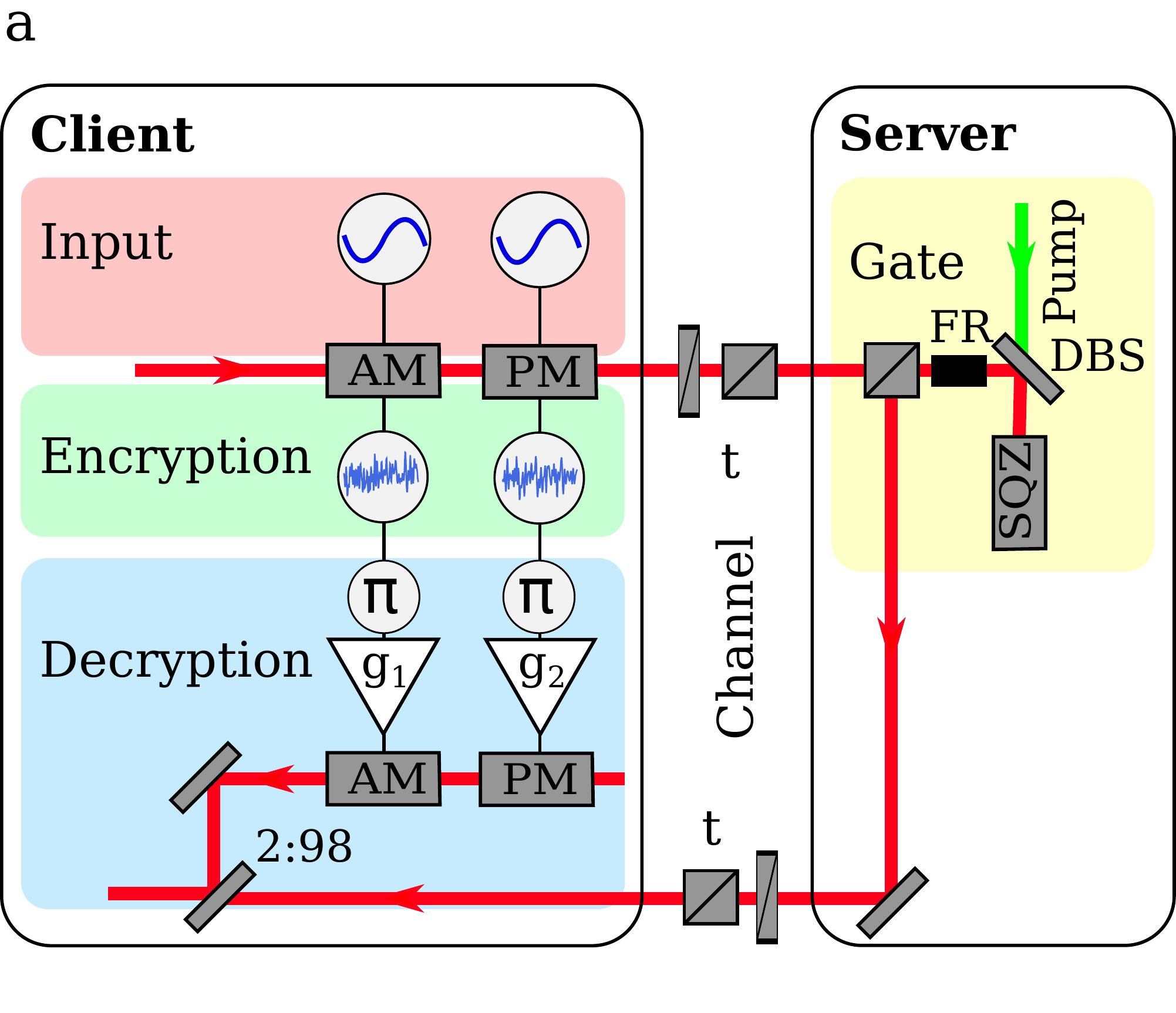}\label{fig:SetupSqueezing}}
    \hspace*{1cm}
    \subfloat{\includegraphics[width=0.5\textwidth]{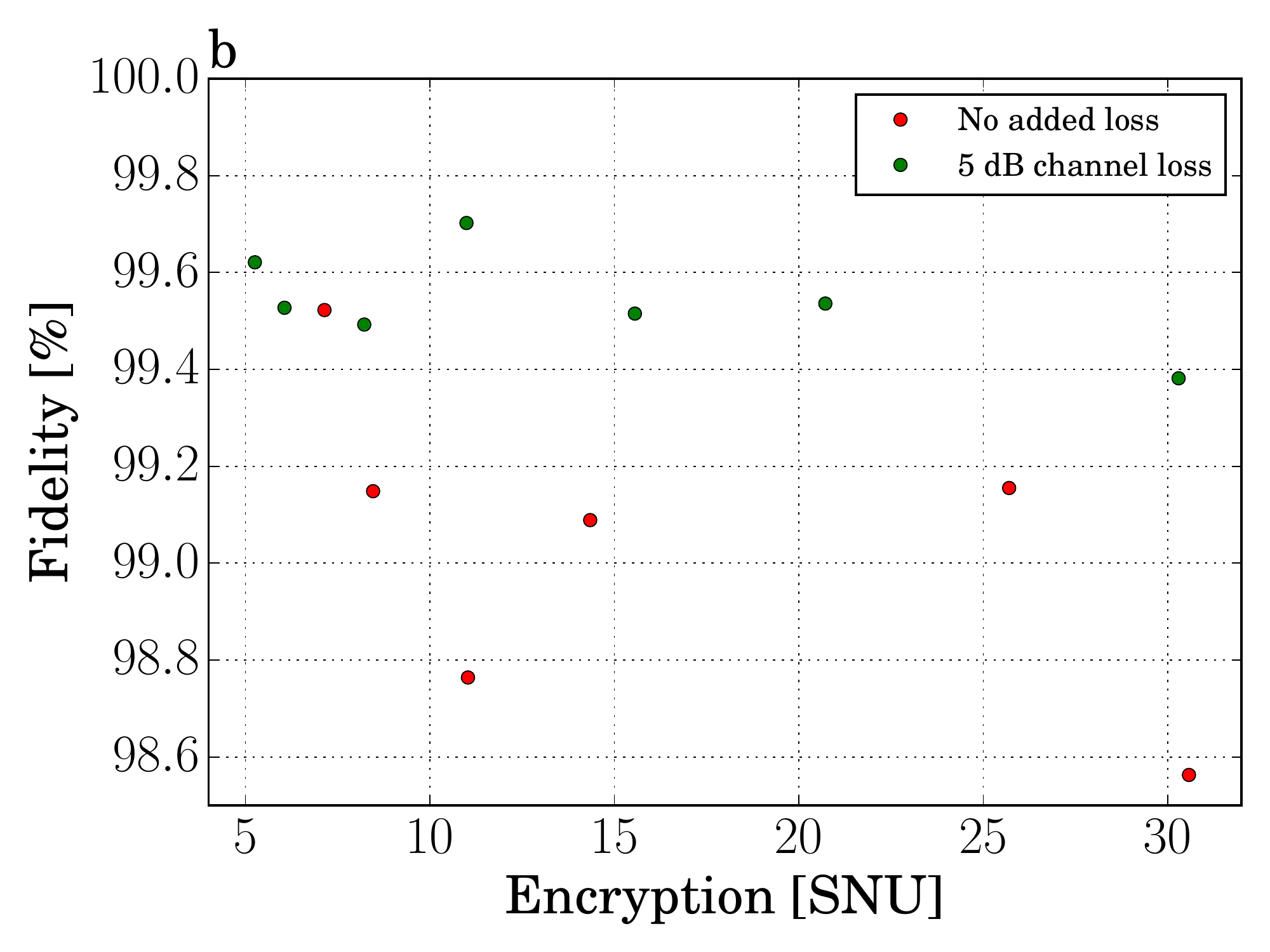}\label{fig:sqz_fidelity}}
    \\
    \subfloat{\includegraphics[width=0.5\textwidth]{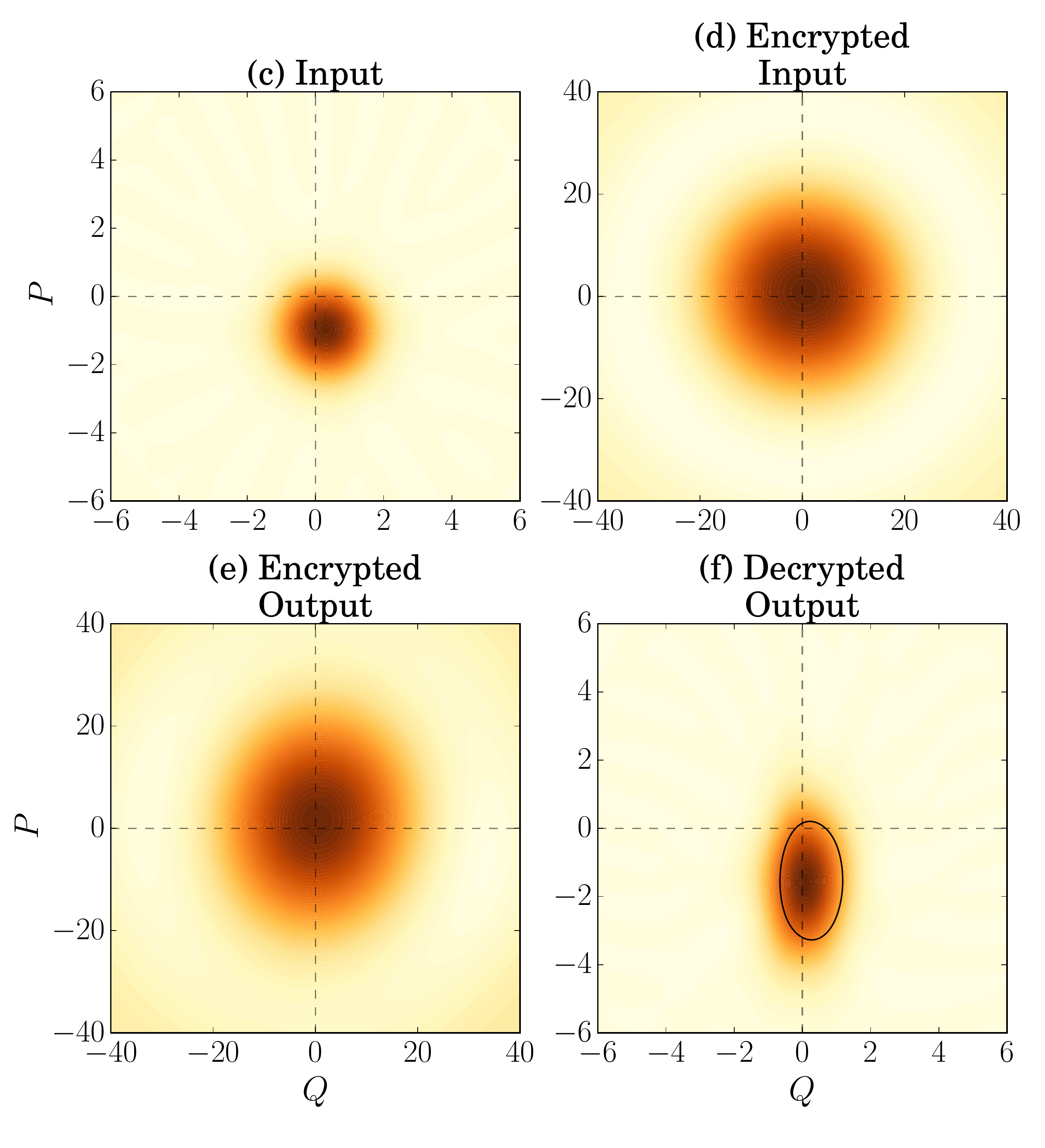}\label{fig:sqz_steps}}
    \subfloat{\includegraphics[width=0.5\textwidth]{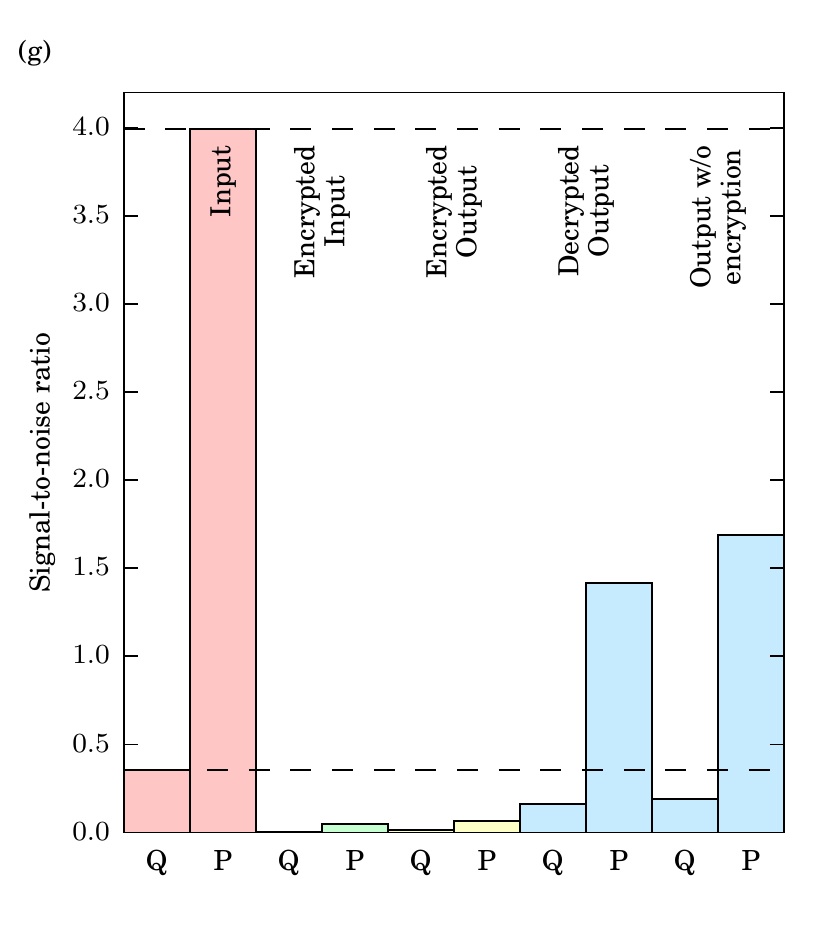}\label{fig:sqz_snr}}
    \caption{\textbf{Server implementing a squeezing gate.} (a) Experimental setup. The squeezing gate was implemented by injecting the input states into a squeezed-light source made of a potassium titanyl phosphate crystal in a linear cavity. After squeezing the output state was sent back to the client who decrypted it by interfering the mode with another beam that was modulated with $\pi$-phase shifted encryption noise at a $2\,\%$ tap-off beam splitter. Due to squeezing the amplitude quadrature, the decryption noise was amplified differently in the two quadratures with gains $g_1$ and $g_2$ depending on the squeezing strength. More details can be found in the supplementary material. (b) The fidelity between the output states of the computation on encrypted and plain-text states with both $100\,\%$ channel transmission and 5\,dB transmission loss corresponding to optical loss in 10\,km fiber at telecom wavelength. Statistical error bars are smaller than the point size. The variation of the fidelities comes mainly from systematic errors in the fine tuning of the decryption and system drifts. (c-f) Reconstructed Wigner functions in phase space for each step in the protocol with a coherent input state and the squeezing gate measured by homodyne detection after each step at a side-band frequency of 10.5\,MHz. The black circle in (f) denotes the FWHM of the ideal squeezing ellipse. (g) shows the signal-to-noise ratios in the Q and P quadratures for the states shown in (c)-(f) as well as for an output state that was not encrypted during computation. The dashed lines indicate the signal-to-noise ratio of an output state using an ideal gate and no channel loss.}
    \label{fig:squeezing}
\end{figure*}

We now turn to the implementation of the squeezing gate which is defined as $S(r)=\exp(r(\hat{a}^2 - \hat{a}^{\dagger 2})/2)$, where $r$ is the squeezing parameter. 
It is directly related to the $U_2(T)$ gate by two additional phase shifts and a suitable transformation between $T$ and $r$~\cite{Miyata2014}, see the supplementary material for a full justification.
Figure~\ref{fig:SetupSqueezing} shows the experimental setup.
In contrast to the implementation of the displacement gates we used a single coherent excitation rather than ensembles of coherent input states.
The squeezed-light source was based on parametric down-conversion in a potassium titanyl phosphate crystal placed in a linear cavity.
In our realization of the gate we directly squeezed the (encrypted) input state in the squeezed-light source. We note that this constitutes the first demonstration of an in-line squeezing transformation of quantum information. Previous demonstrations have relied on off-line squeezed states~\cite{Miyata2014}.
A full Wigner function illustration is presented in Figs.~\ref{fig:squeezing}c-f measured by homodyne detection after each of the four steps. The state after the squeezing gate is shown in Fig.~\ref{fig:squeezing}e, and it is clear that the squeezing operation is hardly visible as a result of the encryption noise. 
Finally, the transformed state is decrypted through displacements at the client, thereby revealing the output state of the server which is displayed in Fig.~\ref{fig:squeezing}f.
The squeezing transformations of the first and second moments of the state are now clearly visible.

We quantify the performance of the squeezing protocol on encrypted data by computing the fidelity between the state retrieved by the client after decrypting the computed input state and the state received by the client when no encryption was used.
This comparison is carried out under the variation of the amount of encryption noise for both no transmission loss and loss corresponding to 10\,km fiber propagation at a telecom wavelength, equivalent to 5 dB transmission loss. This loss is implemented with a half-wave plate and polarizing beam splitter combination.
The results are presented in Fig.~\ref{fig:squeezing}b.
The variation of the fidelities, as was also observed for the displacement gates, comes mainly from systematic errors in the fine tuning of the decryption and system drifts. 
The deviation from unity fidelity is mainly caused by imperfect reconstruction of the Wigner functions, system drifts as well as non-ideal decryption due to a small amount of decorrelation between the encryption and decryption noise.
Despite these imperfections, the fidelity is close to unity and stays above 98.5\,\% for all parameters.
In general the fidelities for the loss case are a bit higher due to the fact that the loss forces the decrypted states closer to the vacuum state.
The measured high fidelities show that performing quantum computing on encrypted states rather than on plain-text states is indeed feasible.

To quantify the performance of the implemented remote gate itself, we display in Fig.~\ref{fig:squeezing}g the SNRs of the $Q$ and $P$ quadratures for the different steps of the protocol.
Ideally, i.e.\ without channel loss, without loss in the gate and with perfect decryption, the squeezing gate will preserve the SNR as indicated by the two dashed lines. 
For the output state we display the SNRs for both with and without encryption, showing a small decrease in SNR if encryption is used.
The remaining reduction of the SNR in comparison to the ideal gate is mainly due to optical loss, i.e.\ channel loss of 5\,dB in each direction and about 2.2\,dB for the gate, including the squeezer and the supply optics.
A more detailed loss analysis can be found in the supplementary material.

\section{Conclusion}

We have developed a continuous-variable protocol for quantum computing on encrypted variables where we required a baseline of only two uses of a quantum channel; one use for the input and another for the output.
We required one additional round of classical communication in each direction and one additional use of the quantum channel to implement a cubic phase gate $U_3(T)$, while Gaussian gates can be implemented with no communication cost.
The client need only be capable of performing displacements to both encrypt and decrypt, except when they perform a $U_3(T)$ gate the client must be capable of implementing a $U_2(A)$ gate as well.
Alternatively, one could run an equivalent entanglement-based version of this protocol which relies on teleportation \cite{Pirandola15}.
To achieve high fidelity teleportation, one could use a hybrid teleportation scheme such as that of Ref.~\cite{Andersen13}.

In this paper, we also studied how to compose gates for encryption and decryption, the effects of transmission and imperfect encryption, an entanglement-based analogue, the use of channel estimation as well as the limitations on squeezing in the cubic phase gate. We have experimentally demonstrated our scheme in performing both displacements and online squeezing operations on an alphabet of coherent states and studied the resulting performance in terms of the fidelity. Finally, to the best of our knowledge, this is the first time quantum computing on encrypted data has been generalized to continuous variables, as well as the first proof-of-principle demonstration of any form (qubit or qumode) of secure delegated quantum computing over a lossy channel. Extending this protocol to long-distances will inevitably require quantum-repeater technologies to preserve the encrypted states \cite{Sangouard11}. We hope our results will help lay the ground work for future theoretical explorations and experimental demonstrations like those for quantum key distribution, such as free-space and field demonstrations.

\bibliography{refs}

\newpage
\appendix
\onecolumngrid
\section{Gate Decryption Operators}
\noindent In this section we first define a set of continuous-variable gates as well as the following conventions.
\begin{align}
[\hat q,\hat p]&=i\\
X(Q)&=\exp(-iQ\hat p)\\
Z(P)&=\exp(iP\hat q)\\
U_k(T)&=\exp(iT\hat q^k)\\
F&=\exp\left[\frac{i\pi}{4}\left(\hat q^2+\hat p^2\right)\right]\\
C_{Z,12}&=\exp(i\hat q_1\otimes\hat q_2)\\
D(\alpha)&=\exp(\alpha \hat a^\dagger - \alpha^* \hat a)\\
D(Q,P)&=e^{-\frac{i}{2}QP}Z(P)X(Q) \text{ for } \alpha=\frac{1}{\sqrt 2}(Q+iP)\\
\ket\psi_{L}&=D(Q,P)\ket\psi
\end{align}
Identities:
\begin{align}
e^Xe^Y&=\exp\{Y+\sum_{n=1}^\infty\frac{1}{n!}[X,Y]_{(n)}\}e^X\\
[X,Y]_{(n)}&=[X,[X,[X,\ldots,[X,Y\underbrace{],\ldots,]]]}_{n-\text{brackets}}
\end{align}
We define a encryption operation, $D(Q,P)$, consisting of a random displacement in phase space.  We consider a universal set of gates $G\in \{X(Q),Z(P),U_2(T),U_3(T),F,C_{Z,12}\}$.  We wish to demonstrate the existence of a correction operator, $C(Q,P,G)$, which depends on the encryption parameters and $G$ such that $C(Q,P,G)GD(Q,P)\ket\psi=G\ket\psi$.
\subsection{$Z(S)$}
\begin{align}
X(Q)Z(S)X^\dagger(Q)&=\exp(-iQ\hat p)\exp(iS\hat q)\exp(iQ\hat p)\\
&=\exp(iS\hat q + [-iQ\hat p,iS\hat q]+\ldots)\\
&=\exp(iS\hat q - iQS)\\
\label{eq:XZ}
\Rightarrow X(Q)Z(S)&=e^{-iQS}Z(S)X(Q)
\end{align}
Correction:
\begin{align}
Z(S)D(Q,P)&=Z(S)e^{-\frac{i}{2}QP}Z(P)X(Q)\\
&=e^{-\frac{i}{2}QP}Z(P)Z(S)X(Q)\\
&=e^{-\frac{i}{2}QP}Z(P)e^{iQS}X(Q)Z(S)\\
&=C^\dagger(Q,P,S)Z(S)\\
C(Q,P,S)&=\exp[i(QP/2-QS)]X(-Q)Z(-P)
\end{align}
\subsection{$X(S)$}
\noindent Correction:
\begin{align}
X(S)D(Q,P)&=X(S)e^{-\frac{i}{2}QP}Z(P)X(Q)\\
&=e^{-\frac{i}{2}QP}e^{-iPS}Z(P)X(S)X(Q) \quad \text{By \eq{XZ}}\\
&=C^\dagger(Q,P,S) X(S)\\
C(Q,P,S)&=\exp[i(QP/2+PS)]X(-Q)Z(-P)
\end{align}
\subsection{$U_2(T)$}
\begin{align}
X(Q)U_2(T)X^\dagger(Q)&=\exp(-iQ\hat p)\exp(iT\hat q^2)\exp(iQ\hat p)\\
&=\exp(iT\hat q^2 + [-iQ\hat p, iT\hat q^2]+\ldots)\\
&=\exp(iT\hat q^2 -2iQT\hat q +\frac{1}{2!}[-iQ\hat p,-2iQT\hat q]+\ldots)\\
&=\exp(iT\hat q^2 -2iQT\hat q +\frac{1}{2}(2iQ^2T)+\ldots)\\
&=\exp(iT\hat q^2 -2iQT\hat q +iQ^2T)\\
\Rightarrow X(Q)U_2(T)&=\exp(iT\hat q^2 -2iQT\hat q +iQ^2T)X(Q)\\
&=e^{iQ^2T}Z(-2QT)U_2(T)X(Q)
\end{align}
Correction:
\begin{align}
U_2(T)D(Q,P)&=U_2(T)e^{-\frac{i}{2}QP}Z(P)X(Q)\\
&=e^{-\frac{i}{2}QP}Z(P)U_2(T)X(Q)\\
&=e^{-\frac{i}{2}QP}Z(P)e^{-iQ^2T}Z(2QT)X(Q)U_2(T)\\
&=C^\dagger(Q,P,T)U_2(T)\\
\Rightarrow C(Q,P,T)&=\exp[i(QP/2+Q^2T)]X(-Q)Z(-2QT)Z(-P)
\end{align}

\subsection{$U_3(T)$}
\begin{align}
X(Q)U_3(T)X^\dagger(Q)&=\exp(-iQ\hat p)\exp(iT\hat q^3)\exp(iQ\hat p)\\
&=\exp(iT\hat q^3 +[-iQ\hat p,iT\hat q^3]+\ldots)\\
&=\exp(iT\hat q^3 -3iQT\hat q^2+\frac{1}{2!}[-iQ\hat p, -3iQT\hat q^2]+\ldots)\\
&=\exp(iT\hat q^3 -3iQT\hat q^2+\frac{1}{2!}(6iQ^2T\hat q)+\frac{1}{3!}[-iQ\hat p,6iQ^2T\hat q]+\ldots)\\
&=\exp(iT\hat q^3 -3iQT\hat q^2+3iQ^2T\hat q+\frac{1}{6}(-6iQ^3T)+\ldots)\\
&=\exp(iT\hat q^3 -3iQT\hat q^2+3iQ^2T\hat q-iQ^3T)\\
\Rightarrow X(Q)U_3(T)&=e^{-iQ^3T}Z(3Q^2T)U_2(-3QT)U_3(T)X(Q)
\end{align}
Correction:
\begin{align}
U_3(T)D(Q,P)&=U_3(T)e^{-\frac{i}{2}QP}Z(P)X(Q)\\
&=e^{-\frac{i}{2}QP}Z(P)U_3(T)X(Q)\\
&=e^{-\frac{i}{2}QP}Z(P) e^{iQ^3T}U_2(3QT)Z(-3Q^2T)X(Q)U_3(T)\\
&=C^\dagger(Q,P,T)U_3(T)\\
C(Q,P,T)&=\exp[i(QP/2-Q^3T)]X(-Q)Z(3Q^2T)U_2(-3QT)Z(-P)
\end{align}
However, $U_2(T)$ does not slide nicely through $F$ so we wish to have the server handle this correction on the fly.
\begin{align}
U_3(T)D(Q,P)&=\exp[i(-QP/2+Q^3T)]U_2(3QT)Z(P-3Q^2T)X(Q)U_3(T)\\
\end{align}
\begin{align}
\Qcircuit @C=1em @R=2em {
\lstick{\ket{\phi}} & \ctrl{1} & \meter & \rstick{\hat p = m_1} \qw \\
\lstick{\ket{0}_p} & \ctrl{-1} & \qw& \rstick{X(m_1)F\ket\phi} \qw
}
\end{align}
Let $\ket\phi\rightarrow F^\dagger U_3(T) D(Q,P)\ket\psi$ so that the output of the teleportation is given by
\begin{align}
&X(m_1)U_3(T) D(Q,P)\ket\psi\\
&=\exp[i(-QP/2+Q^3T)]X(m_1)U_2(3QT)Z(P-3Q^2T)X(Q)U_3(T)\ket\psi.
\end{align}
It is clear that applying the correction operator $U_2(-3QT)X(-m_1)$ to this state would eliminate the appearance of $U_2$, however we cannot divulge the value of $Q$ as this would compromise the security. Consider the circuit given by 
\begin{align}
\Qcircuit @C=1em @R=2em {
\lstick{F^\dagger U_3(T) D(Q,P)\ket\psi} & \ctrl{1} & \meter & \rstick{\hat p = m_1} \qw \\
\lstick{\ket{0}_p} & \ctrl{-1} & \gate{X(-m_1)}& \gate{U_2(A)}& \gate{U_2(B)}& \rstick{Z(P-3Q^2T)X(Q)U_3(T)\ket\psi} \qw,
}
\end{align}
where $A+B=-3QT$.  We can then slide $U_2(A)$ all the way to the left as
\begin{align}
\Qcircuit @C=1em @R=2em {
\lstick{D(Q,P)\ket\psi} &\gate{F^\dagger U_3(T)}& \ctrl{1} & \meter & \rstick{\hat p = m_1} \qw \\
\lstick{U_2(A)Z(Q_2)\ket{0}_p} &\qw& \ctrl{-1} & \gate{Z(2m_1A)}&\gate{X(-m_1)}& \gate{U_2(B)}& \ \qw,
}
\end{align}
where the output is given by $Z(Q_2)Z(2m_1A)Z(P-3Q^2T)X(Q)U_3(T)\ket\psi$ and we also include another $Z$ gate as part of the preparation.  If we choose $A$ at `random' then we are giving the server negligible information about the encryption parameter $Q$ by telling them the value of $B$.  Note that the server will also have to inform the client of $m_1$ obtained so that they can correct for the additional $Z(2m_1A)$ gate, and thus implementing this gate requires a single round of communication.  Furthermore the density operator corresponding to the state $U_2(A)Z(Q_2)\ket{0}_p$ when averaged over $Q_2$ is proportional to the identity, and thus no information about $Q$ can be attained from the state itself. This can be seen from the fact that
\begin{align}
\int dQ_2~ P(Q_2)~Z(Q_2)\ket{0}_p\bra{0}_p Z^\dagger(Q_2)&=\int dQ_2~ P(Q_2)~\ket{Q_2}_p\bra{Q_2}_p\\
\rightarrow \int dQ_2 ~\ket{Q_2}_p\bra{Q_2}_p &\propto \mathbbm{1},
\end{align}
if we choose $Q_2$ according to some Gaussian probability density function, $P(Q_2)$, whose variance tends towards infinity.
\subsection{$C_{Z,12}$}
\begin{align}
C_{Z,12}[X(Q)\otimes \mathbbm{1}]C^\dagger_{Z,12}&=\exp(i\hat q_1\otimes \hat q_2)[\exp(-iQ\hat p_1)\otimes \mathbbm{1}]\exp(-i\hat q_1\otimes \hat q_2)\\
&=\exp(i\hat q_1 \hat q_2)[\exp(-iQ\hat p_1)]\exp(-i\hat q_1 \hat q_2)\quad\text{Tensor structure implicit in subscripts}\\
&=\exp(-iQ\hat p_1 + [i\hat q_1\hat q_2,-iQ\hat p_1]+\ldots)\\
&=\exp(-iQ\hat p_1 + iQ\hat q_2)\\
&=X_1(Q)Z_2(Q)
\end{align}
\begin{align}
C_{Z,12}[X_1(Q_1)\otimes X_2(Q_2)]C^\dagger_{Z,12}&=C_{Z,12}[(X_1(Q_1)\otimes \mathbbm{1}_2)(\mathbbm{1}_1\otimes X_2(Q_2)]C^\dagger_{Z,12}\\
&=C_{Z,12}[(X_1(Q_1)\otimes \mathbbm{1}_2)C^\dagger_{Z,12}C_{Z,12}(\mathbbm{1}_1\otimes X_2(Q_2)]C^\dagger_{Z,12}\\
&=[X_1(Q_1)\otimes Z_2(Q_1)][Z_1(Q_2)\otimes X_2(Q_2)]\\
&=X_1(Q_1)Z_1(Q_2)\otimes Z_2(Q_1)X_2(Q_2)\\
\Rightarrow C_{Z,12}[X_1(Q_1)\otimes X_2(Q_2)]&=[X_1(Q_1)Z_1(Q_2)\otimes Z_2(Q_1)X_2(Q_2)]C_{Z,12}
\end{align}
Correction:
\begin{align}
C_{Z,12}[D_1(Q_1,P_1)\otimes D_2(Q_2,P_2)]&=C_{Z,12}e^{-\frac{i}{2}(Q_1P_1+Q_2P_2)}[Z_1(P_1)X_1(Q_1)\otimes Z_2(P_2)X_2(Q_2)]\\
&=e^{-\frac{i}{2}(Q_1P_1+Q_2P_2)}[Z_1(P_1)\otimes Z_2(P_2)]C_{Z,12}[X_1(Q_1)\otimes X_2(Q_2)]\\
&=e^{-\frac{i}{2}(Q_1P_1+Q_2P_2)}[Z_1(P_1)\otimes Z_2(P_2)][X_1(Q_1)Z_1(Q_2)\otimes Z_2(Q_1)X_2(Q_2)]C_{Z,12}\\
&=e^{-\frac{i}{2}(Q_1P_1+Q_2P_2)}[Z_1(P_1)X_1(Q_1)Z_1(Q_2)\otimes Z_2(P_2)Z_2(Q_1)X_2(Q_2)]C_{Z,12}\\
&=C^\dagger(Q_1,P_1,Q_2,P_2)C_{Z,12}\\
C(Q_1,P_1,Q_2,P_2)&=e^{\frac{i}{2}(Q_1P_1+Q_2P_2)}[Z_1(-Q_2)X_1(-Q_1)Z_1(-P_1)\otimes X_2(-Q_2)Z_2(-Q_1)Z_2(-P_2)]
\end{align}
\subsection{$F$}
\begin{align}
F Z(P) F^\dagger&=X(-P)\\
F X(Q) F^\dagger &= Z(Q)\\
F D(Q,P)&=F e^{-\frac{i}{2}QP}Z(P)X(Q)\\
&=e^{-\frac{i}{2}QP} F Z(P) F^\dagger F X(Q) F^\dagger F\\
&=e^{-\frac{i}{2}QP} X(-P)Z(Q) F\\
&= C^\dagger(Q,P) F\\
C(Q,P)&=e^{\frac{i}{2}QP} Z(-Q)X(P)
\end{align}

\section{Equivalence of the Squeezing Gate}
To demonstrate more than simply shifts in phase space, we implemented a squeezing operation, $S(r)=\exp(r(\hat{a}^2 - \hat{a}^{\dagger 2})/2)$, in the experiment.  To see why this squeezing operation is as challenging as the $U_2(T)$ gate which is required for universal computation, simply notice that it is equivalent up to rotations $U_2(T)=R(\theta)S(r)R(\phi)$ \cite{Miyata2014} where $R(\theta)=\exp(i\theta\hat a^\dagger\hat a)$ is generated by the free Hamiltonian.  To compose gates, these additional phase shifts must be performed  by the server, however to demonstrate a single $U_2(T)$ gate one can notice the following
\begin{align}
C^\dagger(\alpha,T)U_2(T)D(\alpha)\ket \psi&=C^\dagger(\alpha,T)R(\theta)S(r)R(\phi)D(\alpha)\ket \psi\\
&=\tilde C^\dagger(\alpha,T,\theta) S(r) R(\phi)\exp(\alpha \hat a^\dagger-\alpha^*\hat a)R^\dagger(\phi)R(\phi)\ket \psi\\
&=\tilde C^\dagger(\alpha,T,\theta) S(r) \exp(\alpha  e^{i\phi} \hat a^\dagger-\alpha^* e^{-i\phi}\hat a) R(\phi)\ket\psi\\
&=\tilde C^\dagger(\alpha,T,\theta) S(r)  D(\tilde \alpha) \left[ R(\phi) \ket \psi\right].
\end{align}
We have defined a modified decryption operator $\tilde C^\dagger(\alpha,T,\theta)=C^\dagger(\alpha,T) R(\theta)$, reinterpreted the phase reference for the input state $\ket\psi\rightarrow R(\phi) \ket \psi$, and the encryption parameter has simply changed via $\tilde \alpha=\alpha e^{i\phi}$.  A similar trick can be used to shift the rotation $R(\theta)$ through the decryption operator, which is only some displacement, to show that $\tilde C^\dagger(\alpha,T,\theta)=R(\theta)\mathcal C^\dagger(\alpha,T,\theta)$ for some displacement $\mathcal C$.  Thus up to an appropriate redefinition of the initial and final phase references, implementation of a squeezing operation $S(r)$ is sufficient to demonstrate a $U_2(T)$ gate.

\section{Compositions of Gates}
To implement a single gate from the universal set defined above, we have demonstrated that there exists a correction (decryption) operator, $C_G$, that satisfies
\begin{align}
\hat C_G \hat G \hat D(Q,P)\ket\psi&=\hat G\ket\psi,
\end{align}
and similarly for the two-qumode $C_{Z,12}$ gate.  We would like the server to be able to compose gates without needing the client to decrypt (or correct) between applications.  Namely we wish to demonstrate that
\begin{align}
\hat{C'}\hat G_2\hat G_1\hat D(Q,P)\ket\psi=\hat G_2\hat G_1\ket\psi,
\end{align}
or that the client can update their correction operator while only requiring possibly classical communication with the server.  Suppose that
\begin{align}
\hat C_1\hat G_1\hat D(Q,P)\ket\psi=\hat G_1\ket\psi\\
\hat C_2\hat G_2\hat D(Q,P)\ket\psi=\hat G_2\ket\psi,
\end{align}
can we construct a $\hat C'$ from our knowledge of $\hat C_{1,2}$?  Notice that for all of the gates in the universal set the decryption operators, $C$, involve (up to a phase) only products of operators from the set $C\in \{X(Q),Z(P)\}$; aside from the special case of $U_3(T)$ discussed above.  Suppose $G\in \{X(Q),Z(P),U_2(T),U_3(T),F,C_{Z,12}\}$ we would like to show that there exists a modified decryption operator $C'$ such that
\begin{align}
C' G &= G D(Q,P) C
\end{align} 
so that we can \emph{slide} a new correction through the gate as
\begin{align}
C'_1 G_2 G_1 D(Q,P)\ket\psi&= [G_2 D(Q,P) C_1]G_1 D(Q,P)\ket\psi\\
&=G_2 D(Q,P) [C_1G_1 D(Q,P)]\ket\psi\\
&=G_2 D(Q,P) G_1\ket\psi\\
\Rightarrow C_2 C_1' G_2 G_1 D(Q,P)&= C_2 G_2 D(Q,P) G_1\ket\psi\\
&= G_2G_1\ket\psi.
\end{align}
Notice first that gates, $G$, with Hamiltonians involving only powers of $\hat q$ slide trivially through each other.  The only correction operator, in $\{X(Q),Z(P)\}$, that does not involve powers of only $\hat q$ is precisely $X(Q)$.  But we have already shown how to rewrite
\begin{align}
C'G&=GX(Q),
\end{align}
it is easy to insert the extra displacement $D(Q,P)$ between the two terms in the RHS of the above equation since $Z(P),X(Q)$ easily slide through all elements $G$ as illustrated in the above work. Also note for the case where $G=X(Q)$ and $C$ involving powers of $\hat q$ we can apply the same identities by shifting the extra terms to the other side. By recursively applying the identities to swap orderings by applying new corrections we can not only slide the \emph{gate} through the \emph{encryption} operation, we can also slide \emph{corrections} through other \emph{gates}.  In this manner we can build up more complicated gates, with no communication between the client and server for Gaussian gates and one additional round of communication for $U_3$.

\section{Effect of Transmission}
\begin{figure}[hbtp]
\centering
\includegraphics[scale=0.75]{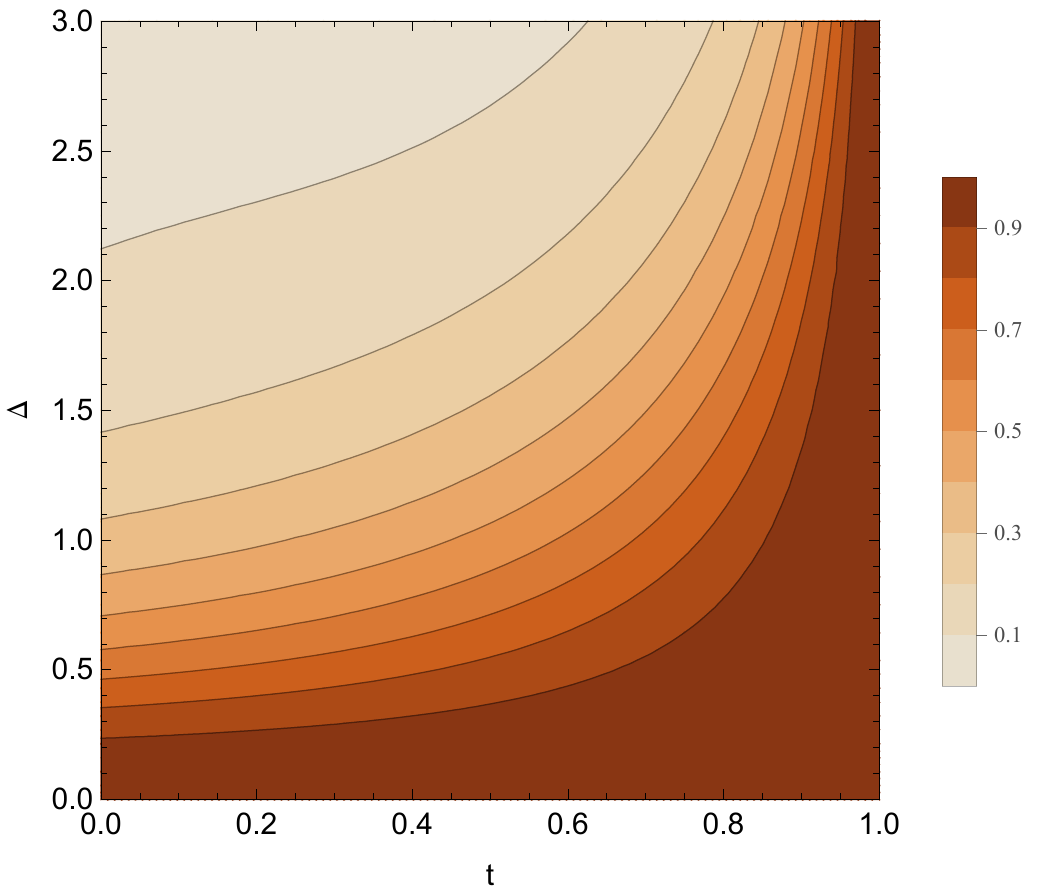}
\caption{Average fidelity between $|\langle \alpha+\beta|t^2\alpha+t\beta\rangle|^2$ for an initial coherent state subject to the protocol with and without loss, which corresponds to a channel with transmission factor $t$.  Both $\alpha$ and $\beta$ are distributed according to Gaussians with zero mean and variance $\Delta^2$.}
\label{fig:fidelity}
\end{figure}

To consider the action of each step of the protocol on our state we move to the Wigner function representation; this is convenient as all operations of interest are linear transformations in the canonical operations $\hat q,\hat p$ and thus are simple transformations of the associated Wigner function.  We model the loss as a beamsplitter $B_{12}(t)$
\begin{align}
B_{12}(t)&=\left(\begin{array}{cc}
\cos\theta & \sin\theta \\ 
-\sin\theta & \cos\theta
\end{array} \right),
\end{align}
where the transmission (reflection) factor is given by $t=\cos\theta$ ($r=\sin\theta$).  This is associated with the input-output relations
\begin{align}
\left(\begin{array}{c}
q_1'\\
q_2'
\end{array}\right)&=B_{12}(t)
\left(\begin{array}{c}
q_1\\
q_2
\end{array}\right),
\end{align}
and similarly for $p_{1,2}$.  When subjecting our mode of interest to loss we inject the vacuum state, $W_{vac}(q,p)=1/\pi\exp(-q^2-p^2)$, in the spare port of the beam splitter.  We can study the effect of loss in comparison to the ideal case by integrating over the loss modes and calculating the fidelity between the resulting Wigner function and the one that would be obtained in the absence of loss.  If the initial state is a coherent state, $\ket\alpha$, then these steps lead to the transformation $\ket\alpha\rightarrow\ket{t^2\alpha+t\beta}$ where $\beta$ is the displacement provided by the server, the general case is presented in~\ref{tab:procedure}.  The symmetric case, where the encryption and the server's displacement are distributed according to a Gaussian probability distribution function with zero mean and variance $\Delta^2$, is plotted in \fig{fidelity}.

\begin{table*}[htp]
\begin{tabular}{l|l|l}
Step & Operation & Transformation \\ 
\hline 
Initial & & $W(q_1,p_1)$ \\ 
Encryption & $D_1(Q,P)$ & $W(q_1-Q,p_1-P)$\\ 
Loss & &  $W(q_1-Q,p_1-P)W_{vac}(q_2,p_2)$\\
& $B_{12}(t)$ & $W(t(q_1-Q)+rq_2,t(p_1-P)+rp_2)\times$ \\
&& $W_{vac}(-r(q_1-Q)+tq_2,-r(p_1-P)+tp_2)$\\
Server & $D_1(A,B)$ & $W(q_1'-A,p_1'-B)W_{vac}(q_2',p_2')$\\
Loss & & $W(q_1'-A,p_1'-B)W_{vac}(q_2',p_2')W_{vac}(q_3,p_3)$\\
& $B_{13}(t)$ & $W(t(q_1'-A)+rq_3,t(p_1'-B)+rp_3)  W_{vac}(q_2',p_2') W_{vac}(-r(q_1'-A)+tq_3,-r(p_1-A')+tp_3)$\\
Decryption & $D_1(-t^2Q,-t^2P)$ & $W(t^2q_1-tA+trq_2+rq_3, t^2p_1-tB+trp_2+rp_3) W_{vac}(q_2',p_2') W_{vac}(q_3',p_3')$ 
\end{tabular}
\caption{The transformations induced by the protocol in the framework of the Wigner function, where we have limited the server to simply applying a displacement.  The calculation for $U_2(T)$ proceeds in a similar fashion, but $U_3(T)$ does not have a linear transformation in terms of the operators $\hat q,\hat p$ and would be more complicated to calculate in this fashion.}
\label{tab:procedure}
\end{table*}
\section{Imperfect Encryption}
To encrypt a state we used the fact that randomly shifting the initial state in phase space results in the maximally mixed state
\begin{align}
X(\ket\psi\bra\psi)=\frac{1}{\pi}\int D(\alpha)\ket\psi\bra\psi D^\dagger(\alpha)d^2\alpha &=\mathbbm{1}.
\end{align}
To see why this is true, notice that $D(\beta)$ provides an irreducible representation, up to a phase, of the Heisenberg-Weyl group.  Furthermore, the operator $X$ commutes with $D(\beta)$ and hence by Schur's Lemma, for unitary groups, $X\propto \mathbbm 1$.  Using the fact that $\ket\psi$ is normalized it is easy to show that the constant of proportionality is $1$, and thus we have the relationship.  From a physical point of view it is not possible to apply a random displacement over $\mathbb R^2$ as this would require infinite energy; from a mathematical perspective it's not clear what a `random' displacement over $\mathbb R^2$ even means.  Instead we consider applying a displacement that is Gaussian distributed with zero mean and variance $\Delta^2$ which follows the mapping
\begin{align}
X(\ket\psi\bra\psi)&=\frac{1}{2\pi\Delta^2}\int D(\alpha)\ket\psi\bra\psi D^\dagger(\alpha)e^{-\frac{|\alpha|^2}{2\Delta^2}}d^2\alpha.
\end{align}
The finite width of the Gaussian will lead to this state differing from the maximally mixed state; in general the resulting state will depend on the initial state.  A simple figure of merit we can study is the purity of the state $\text{Tr}\rho^2$, in discrete variable systems the purity is always greater than $1/d$ but in continuous variable systems the purity can be zero, as is the case for the maximally mixed state.  We can find the purity of an arbitrary initial state, $\rho\rightarrow W(\alpha)$, subject to the map $X(\ket\psi\bra\psi)$ by working in the Wigner function formalism
\begin{align}
W(\alpha)\rightarrow W(\alpha')&=\frac{1}{2\pi\Delta^2}\int W(\alpha-\beta)e^{-\frac{|\beta|^2}{2\Delta^2}}d^2\beta\\
\text{Tr}\rho^2&=2\pi\int W(\alpha')^2d^2\alpha'.
\end{align}
Note that this transformation is also just the convolution of our initial Wigner function with a Gaussian probability distribution.  We can then use the purity to characterize how well our encryption operation simulates the ideal encryption procedure, given a particular class of input states.  Consider the case where our initial state is also a coherent state, given by $\ket\alpha$
\begin{align}
X(\ket\alpha\bra\alpha)&=\frac{1}{2\pi\Delta^2}\int D(\beta)\ket\alpha\bra\alpha D^\dagger(\beta)e^{-\frac{|\beta|^2}{2\Delta^2}}d^2\beta\\
&=\frac{1}{2\pi\Delta^2}\int \ket{\alpha+\beta}\bra{\alpha+\beta} e^{-\frac{|\beta|^2}{2\Delta^2}}d^2\beta\\
&=\frac{1}{2\pi\Delta^2}\int \ket{\gamma}\bra{\gamma} e^{-\frac{|\gamma-\alpha|^2}{2\Delta^2}}d^2\gamma.
\end{align}
Note that a thermal state $\rho_{th}(\bar n)$ of average excitation number $\bar n$ has Glauber-Sudarshan representation given by
\begin{align}
\rho_{th}(\bar n)&=\frac{1}{\pi\bar n}\int\ket\alpha\bra\alpha e^{-\frac{|\alpha|^2}{\bar n}}d^2\alpha.
\end{align}
Thus we have that $X(\ket\alpha\bra\alpha)$ corresponds to a thermal state with $\bar n=2\Delta^2$ which is displaced by an amount $\alpha$.  For a fixed amount of energy, thermal states maximize their Von Neumann entropy $S(\rho)=-\text{Tr}(\rho\log\rho)$ which is also related to purity; $S(\rho)$ is maximized by the maximally mixed state and is zero for pure states.  The purity of $\rho_{th}(\bar n)$ can be calculated as $\text{Tr}(\rho_{th}(\bar n)^2)=1/(1+2\bar n)$ so that the purity of $X(\ket\alpha\bra\alpha)$ is given by $1/(1+4\Delta^2)$.

\begin{figure}[hbtp]
\centering
\includegraphics[scale=1]{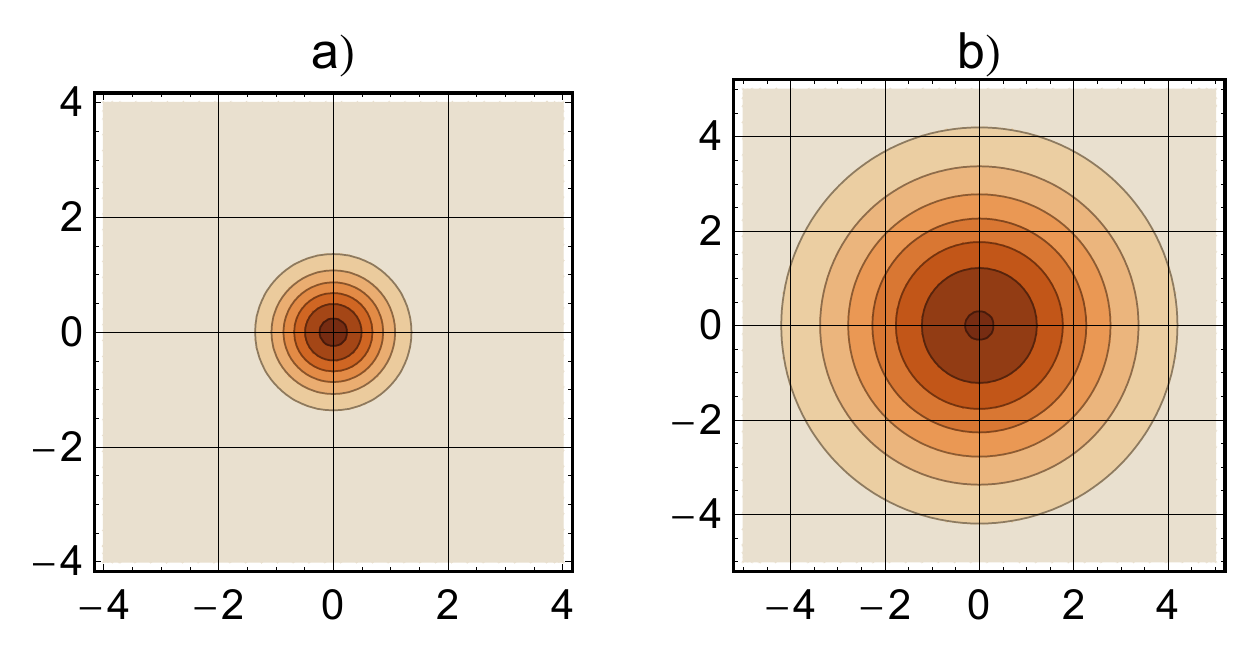}
\caption{(a) The Wigner function corresponding to the vacuum state. (b) The Wigner function after the vacuum is encrypted where $\Delta=2$; this corresponds to a thermal state with $\bar n=8$.}
\label{fig:coherent}
\end{figure}

\begin{figure}[hbtp]
\centering
\includegraphics[scale=1]{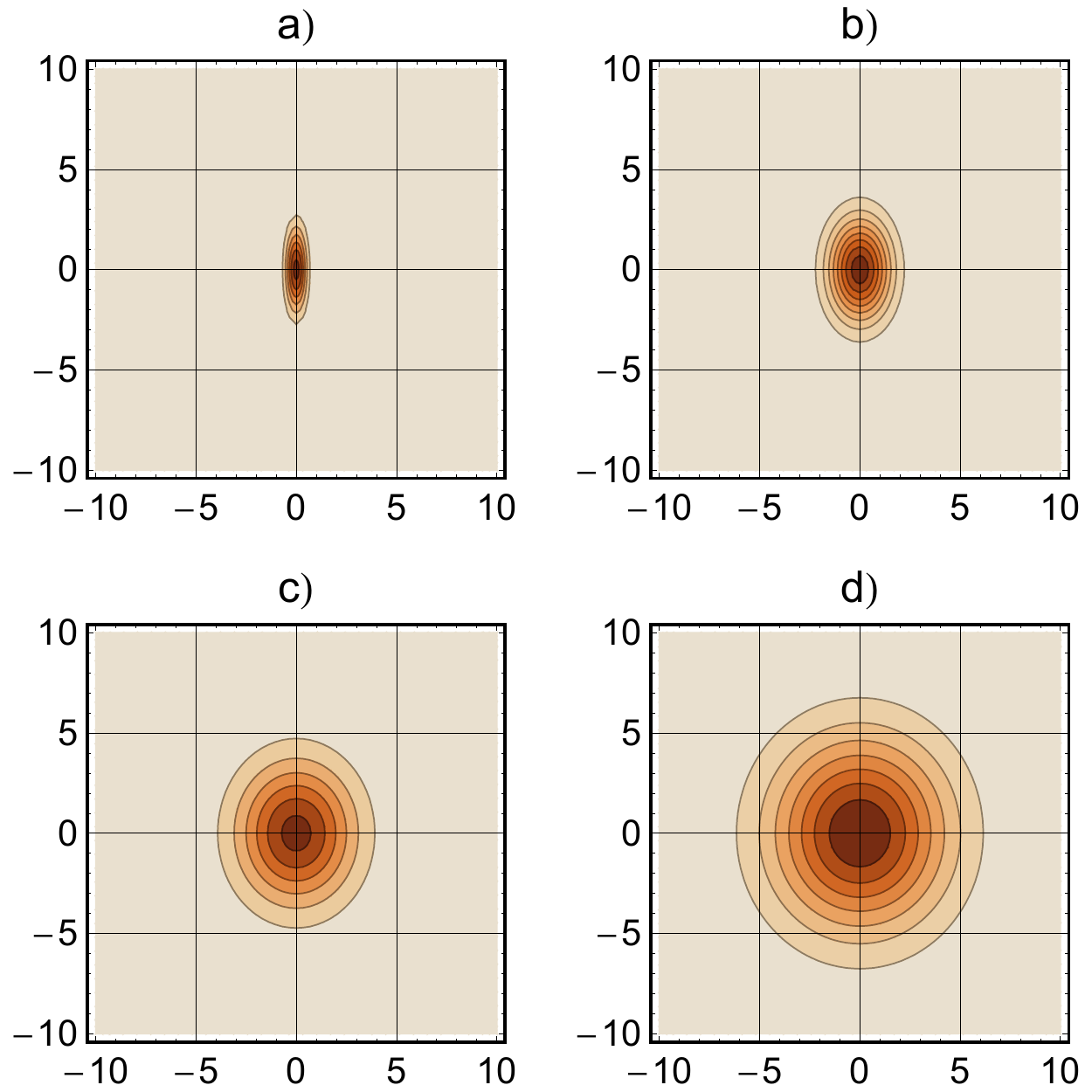}
\caption{(a) The Wigner function for a squeezed vacuum state with squeezing parameter $r=\ln2$. (b,c,d) The Wigner function after the squeezed state is encrypted with parameter $\Delta=1,2,3$ respectively.  Notice that for small $\Delta$ the encrypted state is asymmetric, which gives us some information about the initial state, but for higher values the plot more closely resembles a thermal state.  The purity of the resulting states (a-d) are $\approx 1,0.27,0.10,0.05$ respectively.}
\label{fig:squeezed}
\end{figure}

In general one can use the Wigner function to evaluate the action of the encryption and to determine how mixed the resulting state is. In \fig{coherent} we show the action of imperfect encryption on the vacuum state and in \fig{squeezed} we show the same for a squeezed vacuum state. In general, we want the Gaussian envelope to be large enough to smear out any details about the set of possible initial states.

\section{Entanglement-Based Analogy}
One can exploit the well-known equivalence between preparing a coherent state, chosen according to a Gaussian distribution, and performing Heterodyne detection on half of an EPR pair to show an analogous entanglement-based encryption scheme.  In conventional CV teleportation \cite{Braunstein98}, where the parties share a perfectly squeezed EPR pair, the final correction operator is precisely a displacement.  Using the principle of deferred measurement, where we simply wait to perform the two homodyne measurements until after the state has been returned from the server, it is clear that the server could not have known of the measurement outcomes and hence the encryption parameters.  This analogy works only in the limit of infinite squeezing; in the realistic scenario of finite squeezing there is excess noise introduced onto the mode of interest and the entanglement-based scheme is not a perfect analogy.  However, in the limit of large squeezing, or equivalently a Gaussian distribution with large variance, the excess noise tends to zero and this is a good approximation to the actual implementation.

One can also show that the interactive $U_3(T)$ gate has an entanglement-based equivalent.  First note that,

\begin{align}
U_{2,b}(A)\ket{EPR}&\propto U_{2,b}(A)\int dq \ket q_a\ket q_b\\
&\propto \int dq e^{iA q^2} \int dp e^{-iqp} \ket q_a\ket p_b.
\end{align}
If we now measure the second mode and obtain an outcome $P$ we have the following,
\begin{align}
&\rightarrow \int dq e^{iA q^2} e^{-iqP}\ket q_a\\
&= U_{2,a}(A) \int dq e^{-iqP}\ket q_a\\
&\propto U_{2}(A) Z(-P)\ket 0_p.
\end{align}
Thus we can use the above procedure to modify the implementation of $U_3(T)$ and defer the final measurement to construct an entanglement-based version.  Explicitly, the client can implement $U_2(A)$ on half of an EPR pair, shared with the server, before measuring the $\hat p$ quadrature; this will have the same effect as the previous scheme.  As before, this holds in an exact sense only in the unphysical limit of perfect squeezing.
\section{Channel Estimation}
When accounting for loss, as seen in~\ref{tab:procedure}, we require the displacement $D(-t^2Q,-t^2P)$ as part of the correction.  To perform this displacement we must know the transmission coefficient $t$, however we could forgo this stipulation and assume that $t\approx 1$, but doing so will result in a reduced fidelity.  In order to partially correct for loss it is necessary to do parameter estimation on the channel.  This can be done, for example, by periodically sending random coherent states and asking the server to apply some random displacement, after which one can measure the state in order to learn information about the loss parameter.  Generally, there will also be additional noise that one can not correct for as well, but if desired one can estimate the magnitude of this noise by sending an appropriate ensemble of initial states.  By doing this frequently enough one can ensure that the channel is not varying appreciably with time, and if so to adjust the necessary correction.  

This process ensures that the server is not maliciously altering the channel parameters, in fact we can take this notion one step further and check that the server is indeed performing the desired displacement thus verifying the $X,Z$ operations in our full calculation.  In the original proposal for blind quantum computation, the ability to verify that the server is performing the desired calculation arises from the capability of hiding the gates that one is performing.  Through doing so one is able to periodically perform a check by doing an easy computation of which one already knows the answer.  Since the server is unable to discriminate this check from the actual computation it is unable to pass the check without operating faithfully.  In our case the server is fully aware of all gates being performed and could for example perform all operations except $U_3(T)$ as intended.  Fortunately, there is a large class of desirable computations for which there exist efficient classical verification methods and so one is able to detect an incorrect answer with classical post-processing.

\begin{figure}[hbtp]
\centering
\includegraphics[scale=0.90]{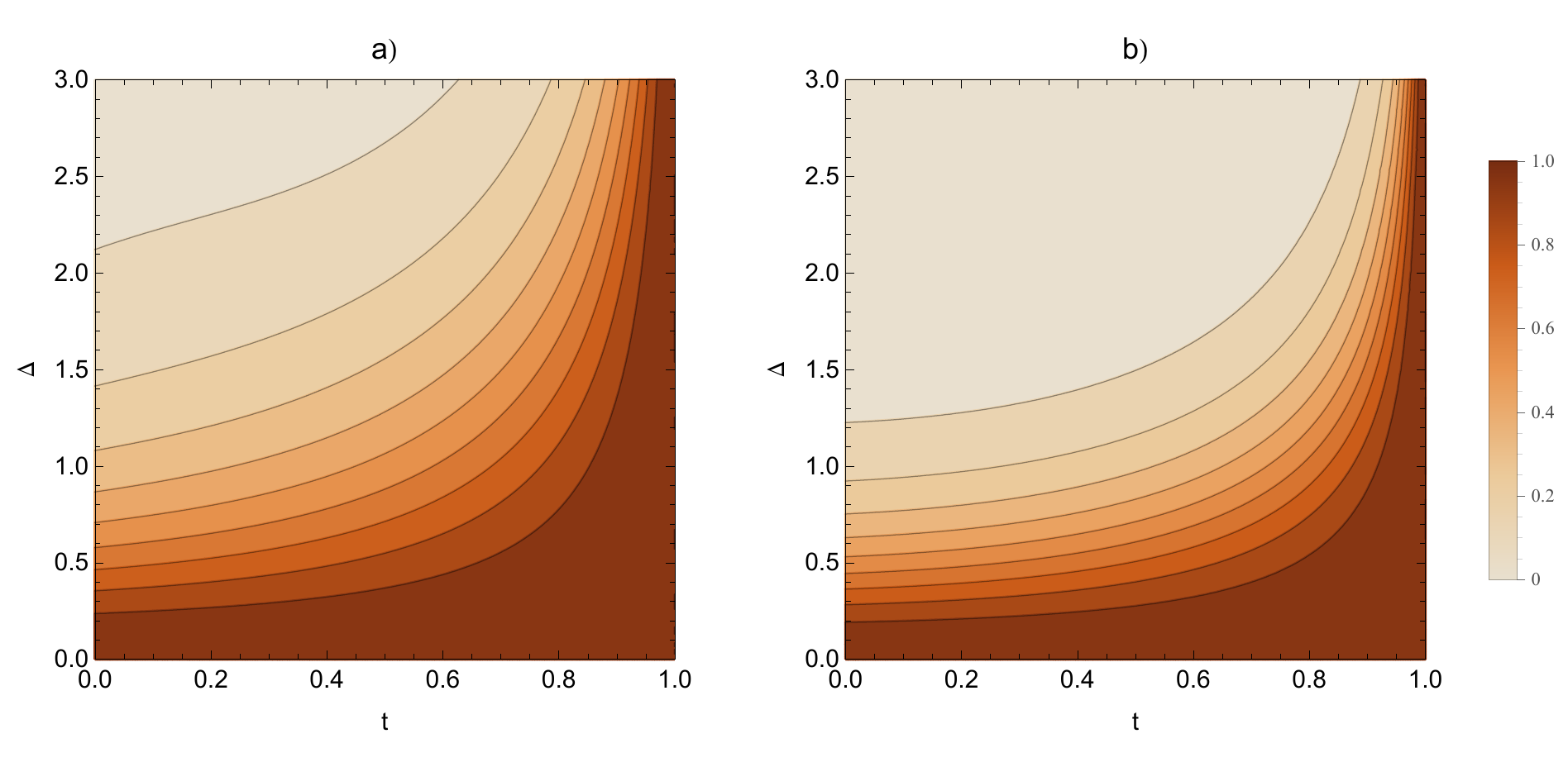}
\caption{(a) Average fidelity between $|\langle \alpha+\beta|t^2\alpha+t\beta\rangle|^2$ for an initial coherent state subject to the protocol with and without loss, which corresponds to a channel with transmission factor $t$.  (b) The same plot as in part (a) but where the final displacement does not take into account the channel loss $t$, given by $|\langle \alpha+\beta|t^2\alpha+t\beta\rangle+(t^2-1)\gamma|^2$, where $\gamma$ is the encryption parameter.  All of $\alpha,\beta$ and $\gamma$ are distributed according to Gaussians with zero mean and variance $\Delta^2$.}
\label{fig:channelestimate}
\end{figure}

We show the advantage of accounting for the loss in the channel by comparing the fidelities to the ideal state both with and without taking into account the loss parameter in \fig{channelestimate}.
\section{Limitations on $U_2(T)$}
Suppose we wish to decompose $U_2(X_\text{tot})=U_2(X_\text{in})U_2(X_\text{enc})$ as the sum of two parts $X_\text{tot}=X_\text{in}+X_\text{enc}$, as in the case where we split the operator $U_2(-3QT)$ between the client and server in order to implement $U_3(T)$. Furthermore, assume that we do not want the knowledge of $X_\text{enc}$ to reveal the value of $X_\text{tot}$.  To attempt to hide the value of $X_\text{tot}$, we choose $X_\text{in}$ to be a random variable of zero mean and variance $V_\text{in}$.  One method of characterizing the amount of information that a random variable carries about a parameter upon which its probability distribution depends on is with the Fisher information, given by
\begin{align}
\mathcal I_X(\theta)&=\mathbb E\left[\left(\frac{\partial}{\partial\theta}\log P(X;\theta)\right)^2\Bigg | \theta\right].
\end{align}
Explicitly, in the case where $X_\text{tot}=-3QT$, the Fisher information for $X_\text{enc}$ has the property that $\mathcal I(Q)\propto T^2/V_\text{in}$.  Notice that as $V_\text{in}\rightarrow 0$ the Fisher information increases without bound indicating that $X_\text{enc}$ allows one to make a very good estimate of the encryption parameter $Q$.  However, when $V_\text{in}\rightarrow \infty$ then Fisher information approaches zero and the server learns no information of the encryption parameter.

\section{Measuring encryption efficiency}

\begin{figure}
\includegraphics[width=0.4\textwidth]{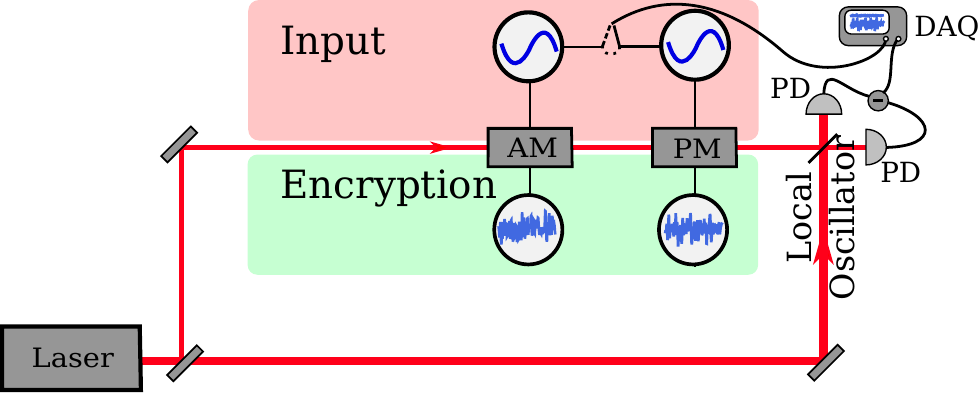}
\caption{A schematic representation of the encryption efficiency experiment. PD: Photo diode. DAQ: Data acquisition.}
\label{fig:SetupEncryptionEffectivenessSupp}
\end{figure}

A shot-noise limited laser at 1064 nm generated a highly coherent beam, which was split into two. One part was designated as a local oscillator (LO), the other as the signal beam. A sketch of the setup is shown in Fig.~\ref{fig:SetupEncryptionEffectivenessSupp}.

The local oscillator was sent directly to the detection stage, while the signal beam was passed through a set of optical modulators, to be modulated in the phase and amplitude quadratures. The electronic input to the modulators in either quadrature was an addition of two signals, one representing the alphabet of the different input states of the client, the other the client's encryption noise. The noise was determined to be white within the measurement bandwidth of 1 MHz. The alphabet was recorded by the data acquisition to establish the size of the correlations. The homodyne detector output was demodulated at 10.5 MHz, with a 1 MHz lowpass filter to set the measurement bandwidth. The sampling rate was 5 MHz, with 14 bit resolution. The variance of the Gaussian distributed input state alphabet was $V_{\text{in}} = 0.6$ SNU. From the measured data we determined the variances and covariances which give an estimate for the mutual information by the formula,

\begin{equation}
I(\text{server}_{\text{enc}}:\text{client}_{\text{in}}) = \dfrac{1}{2} \log_2 \left(\dfrac{V_{\text{in}}}{V_{\text{in}} - \frac{C^2}{V}} \right),
\end{equation}

\noindent where $C$ is the covariance between the recorded alphabet and the signal, and $V = V_\text{in} + V_\text{enc}$, i.e. the variance of the signal and the encryption.

\subsection{Displacement gates}

\begin{figure}
\includegraphics[width=0.4\textwidth]{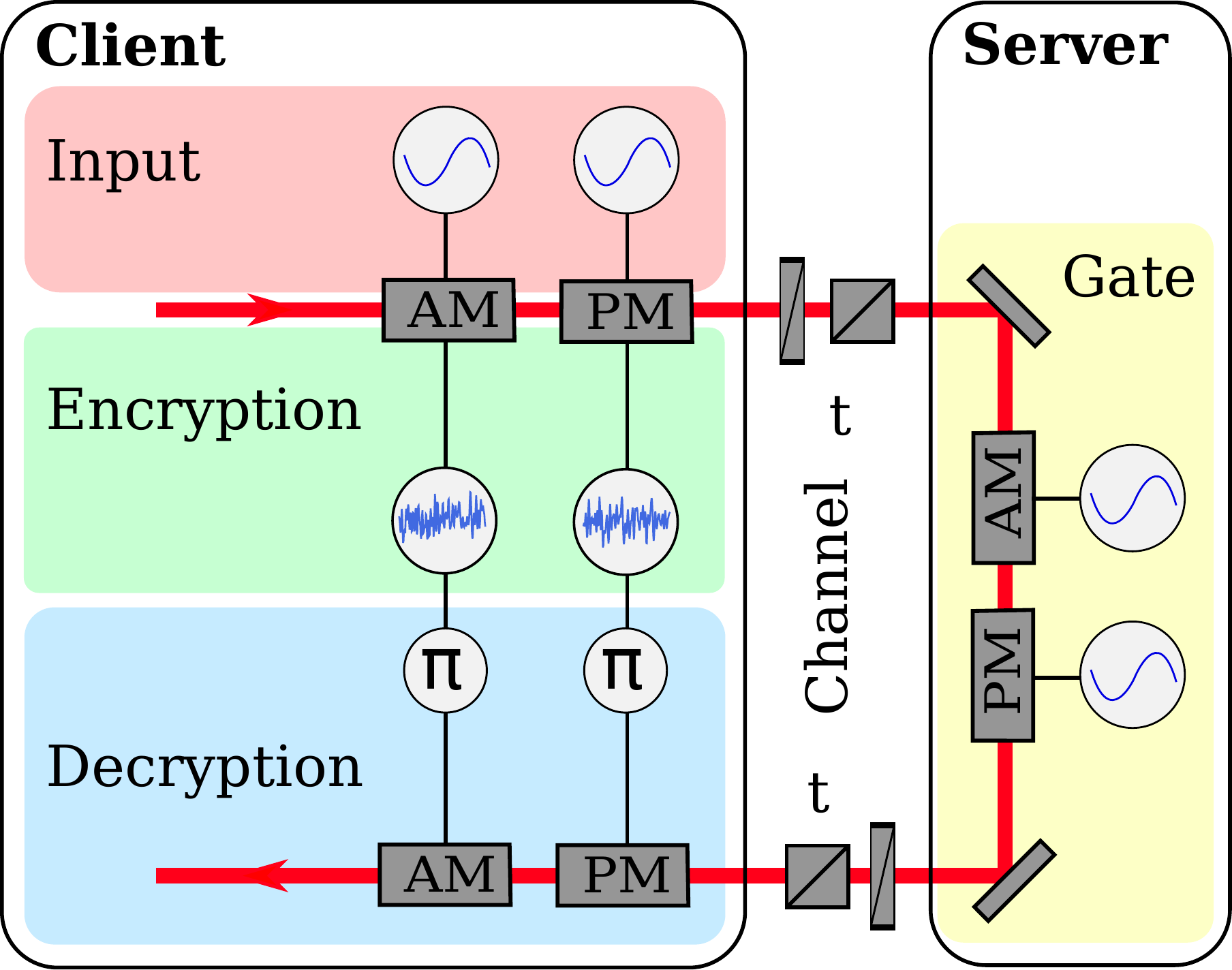}
\caption{A schematic representation of the displacement gate experiment.}
\label{fig:SetupDisplacementsSupp}
\end{figure}

A shot-noise limited laser at 1064 nm generated a highly coherent beam, which was split into two. One part was designated as a local oscillator (LO), the other as the signal beam. A sketch of the setup is shown in Fig.~\ref{fig:SetupDisplacementsSupp}.

The local oscillator was sent directly to the detection stage, while the signal beam was passed through a set of optical modulators, to be modulated in the phase and amplitude quadratures. The electronic input to the modulators in either quadrature was an addition of two noise signals, one representing the alphabet of the different input states of the client, the other the client's encryption noise. The noise signals were determined to be white within the measurement bandwidth of 1 MHz. The state then passed through a half-wave plate and polarizing beam splitter combination which simulated an attenuation $t$ from the client to the server.

Then the state was further displaced by the server, with another signal representing the alphabet of a simple linear gate. Like the previous modulations the displacement signal was also white within the measurement bandwidth. The state was then attenuated with another polarizing beam splitter and a half-wave plate, with a setting identical to the first. This was done to simulate a state experiencing the same loss on the return trip from the server to the client. Following this final loss simulation, the state was modulated again by the client, to eliminate the encryption noise. This elimination was possible down to less than 0.2 SNU in both quadratures. This was in part made possible due to a custom-made noise generator with extremely well correlated outputs at the relevant frequencies. To obtain the right phase between the noise modulation and the cancelling modulation, a controllable phase delay between the modulations was also introduced, by using a DB64 Coax Delay Box from Stanford Research Systems from the noise generator to the modulator. The state was then interfered with the local oscillator at a 50/50 beam splitter, with the output modes subsequently detected by PIN photo diodes with a quantum efficiency of 90 \%. The visibility of the interference was 92 \%, due to the distortion of the spatial beam profile by the many modulators. Obtaining this visibility was made easier by introducing a cavity to which both the signal and the local oscillator were matched, to ensure efficient overlap despite the spatial distortions.

Two types of measurements were made. The first type was where the relative phase of the local oscillator with the signal was continuously scanned by a slow function generator at 5 Hz. The outputs of the two photo detectors were subtracted and mixed down to DC using a strong electronic local oscillator at 10.5 MHz. The output of the mixer was amplified and low-pass filtered at 1 MHz before being sent to a data acquisition card (DAC) where the voltages were sampled with a rate of 5 MHz. Digitally a 10 kHz high-pass filter was implemented to dampen a strong 50 Hz modulation originating from the AC line, which was detrimental to the quality of the state reconstruction. The measurements were used to reconstruct the density matrices and Wigner functions of the input and output states using a maximum likelihood algorithm~\cite{Lvovsky2009} and the Python module QuTiP \cite{Johansson2013}.

\begin{figure}[hbtp]
\begin{center}
    \subfloat{\includegraphics[width=8.5cm]{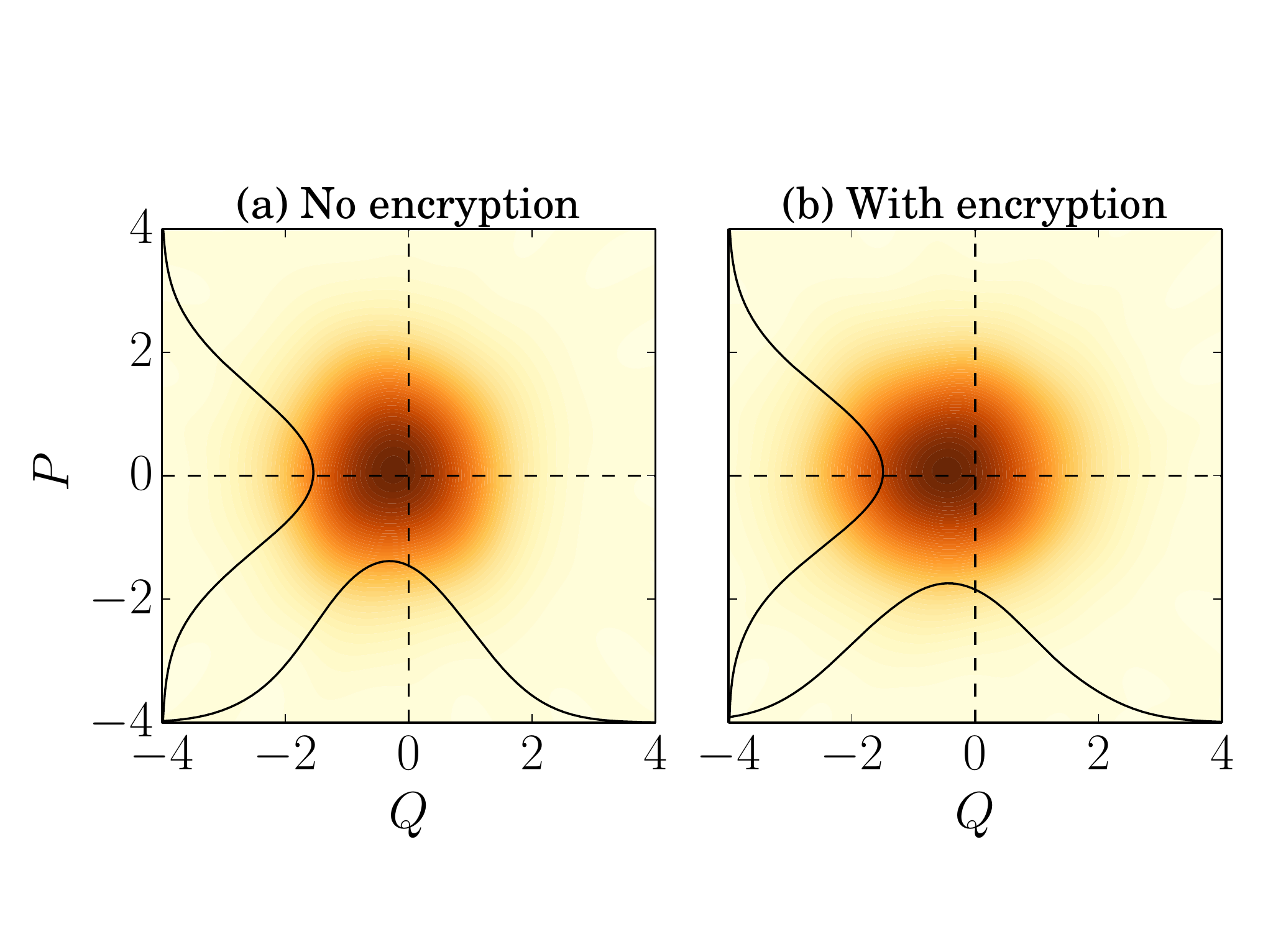}}\\
    	\vspace*{-1cm}
    \subfloat{\includegraphics[width=8.5cm]{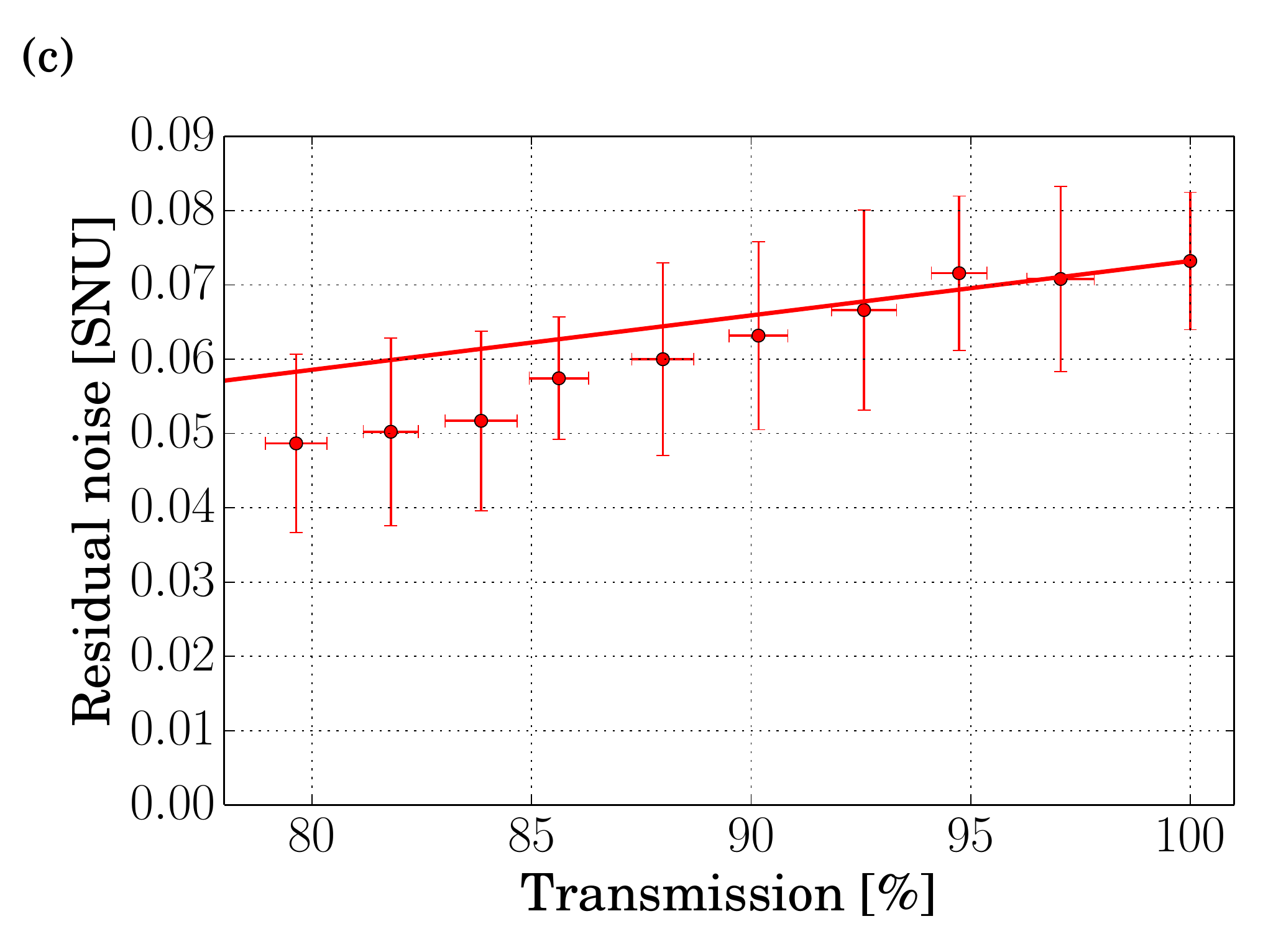}}
    
\caption{\textbf{The effect of encryption and decryption on the computed data.} (a,b) Reconstructed Wigner functions of the ensembles of coherent states representing the output states, i.e.\ the input state ($V_\text{in}=0.3$\,SNU) after the gate operation ($V_\text{gate}=0.6$\,SNU). (a) Without encrypting the data and (b) with encryption ($V_\text{enc}=31$\,SNU). (c) Plot of the residual noise of the decryption process as a function of transmission. The solid line is a theory curve of the residual noise versus transmission calculated by starting with the measured residual noise value of the ideal channel ($T=100\,\%$). The data was recorded with a spectrum analyser measuring zero-span around $10.5$\,MHz with a resolution bandwidth of $300$\,kHz and a video bandwidth of $30$\,Hz. The size of the error bars is due to the standard deviation of the $1000$ recorded points.}
\label{fig:DispGateResultsSupp}
\end{center}
\end{figure}

The first measurements were used to estimate the fidelity of the decryption operation relative to an unencrypted state going through the gate and the channel, applying the definition of fidelity \cite{Weedbrook12},

\begin{equation} \label{eq:Fidelity}
F = \text{Tr} \left(\sqrt{\sqrt{\rho_0} \rho_1 \sqrt{\rho_0}} \right),
\end{equation}

to the reconstructed density matrices. The different Wigner functions are shown in Fig.~\ref{fig:DispGateResultsSupp}a and b.

The second type of measurement had the relative phase locked such that the phase quadrature was measured. Here the subtracted photo detector signals were sent to a spectrum analyzer (SA), which monitored the signal strength using a zero span trace at 10.5 MHz. This was done to easily monitor the variance of the applied Gaussian noise modulations. Though this variance was only monitored in the phase quadrature, the state reconstruction was done on-line to ensure generation of states with approximately symmetric variances. The estimation of the residual noise can be seen in Figure \ref{fig:DispGateResultsSupp}c.

\subsection{Squeezing gate}

\begin{figure}
\includegraphics[width=0.4\textwidth]{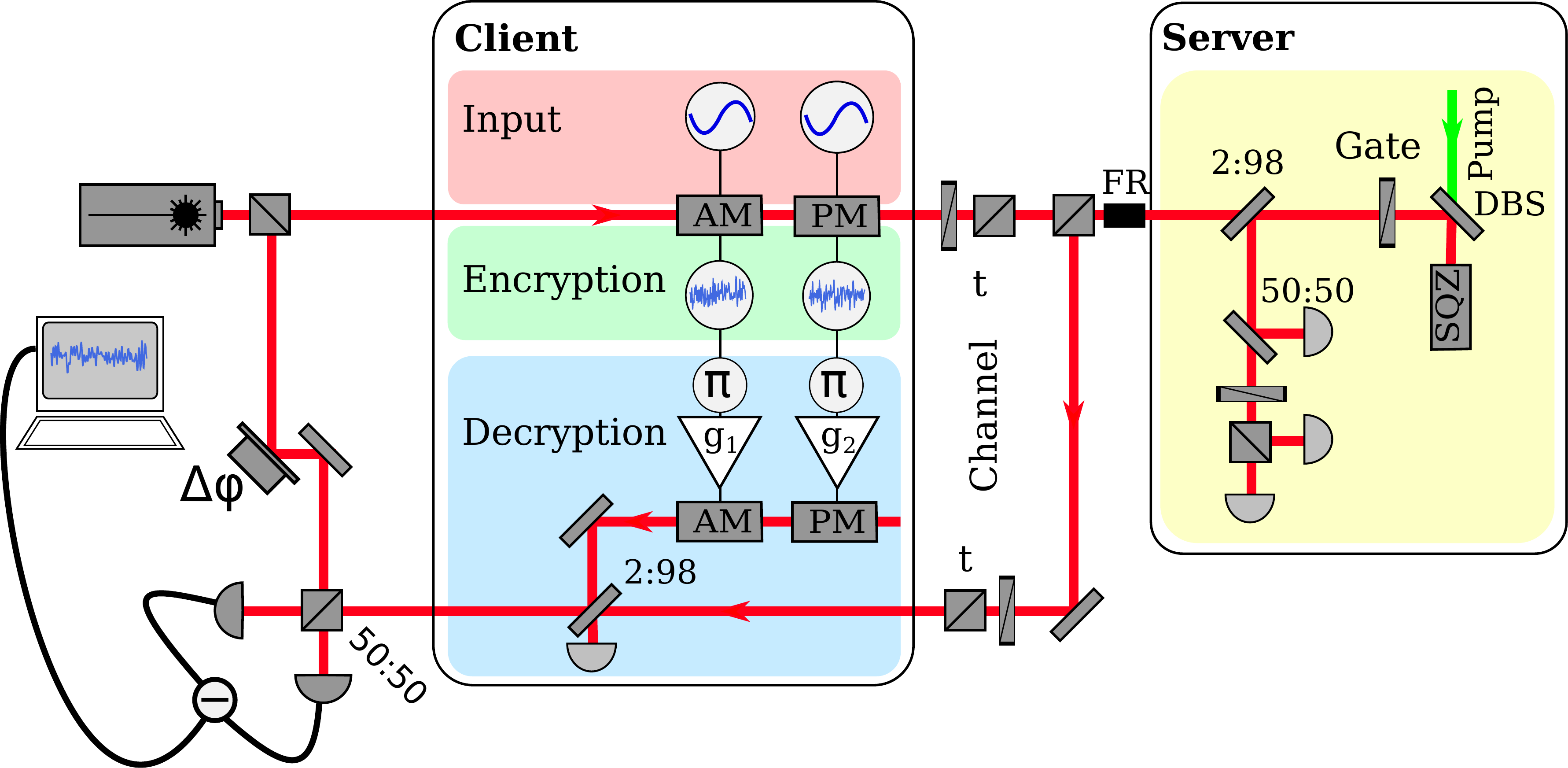}
\caption{A schematic representation of the squeezing gate experiment.}
\label{fig:SetupSqueezingSupp}
\end{figure}

A shot-noise limited laser at 1064 nm generated a highly coherent beam, which was split into three. One part was designated as a local oscillator (LO), the second as the signal beam, and the third as the decryption beam. A sketch of the setup is shown in Fig.~\ref{fig:SetupSqueezingSupp}.

The signal beam was modulated with a single displacement in $Q$ and $P$, and in addition a noise modulation generated by a noise generator. 
Another output of each of the noise generators, highly correlated with the first, was connected to the corresponding modulators in the decryption path.
The signal beam then encountered a half-wave plate and polarizing beam splitter combination to simulate the channel, and was then forwarded through a Faraday rotator and another half-wave plate to ensure that the light was mostly in the s-polarization before entering the squeezing cavity.
The light entered the resonant linear semi-monolithic cavity containing the $1 \times 2 \times 10\,\text{mm}^3$ periodically poled potassium titanyl phosphate crystal pumped with a 532\,nm pump beam of 7\,mW through the coupling mirror.
The outer crystal face was coated to have a high reflectivity of 99.95\,\% for both wavelengths, while a curved mirror with a radius of curvature of 20\,mm, a reflectivity of 90\,\% for the fundamental wavelength at 1064\,nm and a reflectivity of 20\,\% for the pump wavelength served as a piezo-tunable coupling mirror.
The crystal was kept at a phase matching temperature of $\SI{36.2}{\celsius}$ , and most of the signal returned from where it entered, towards the Faraday rotator, but having been squeezed by performing round trips in the cavity.
Before hitting the Faraday rotator, a beam sampler redirected 2\,\% of the light for generating an error signal for cavity and pump phase locking.
Because the squeezing cavity was birefringent, the small p-polarization component was used to generate an error signal with the H\"ansch-Couillaud locking technique~\cite{Hansch1980}.
The phase of the pump beam with respect to the signal beam was locked using a phase modulation of the signal beam at 36.7\,MHz generated using the phase modulator used for encrypted input state generation.

\begin{figure}
\includegraphics[width=0.4\textwidth]{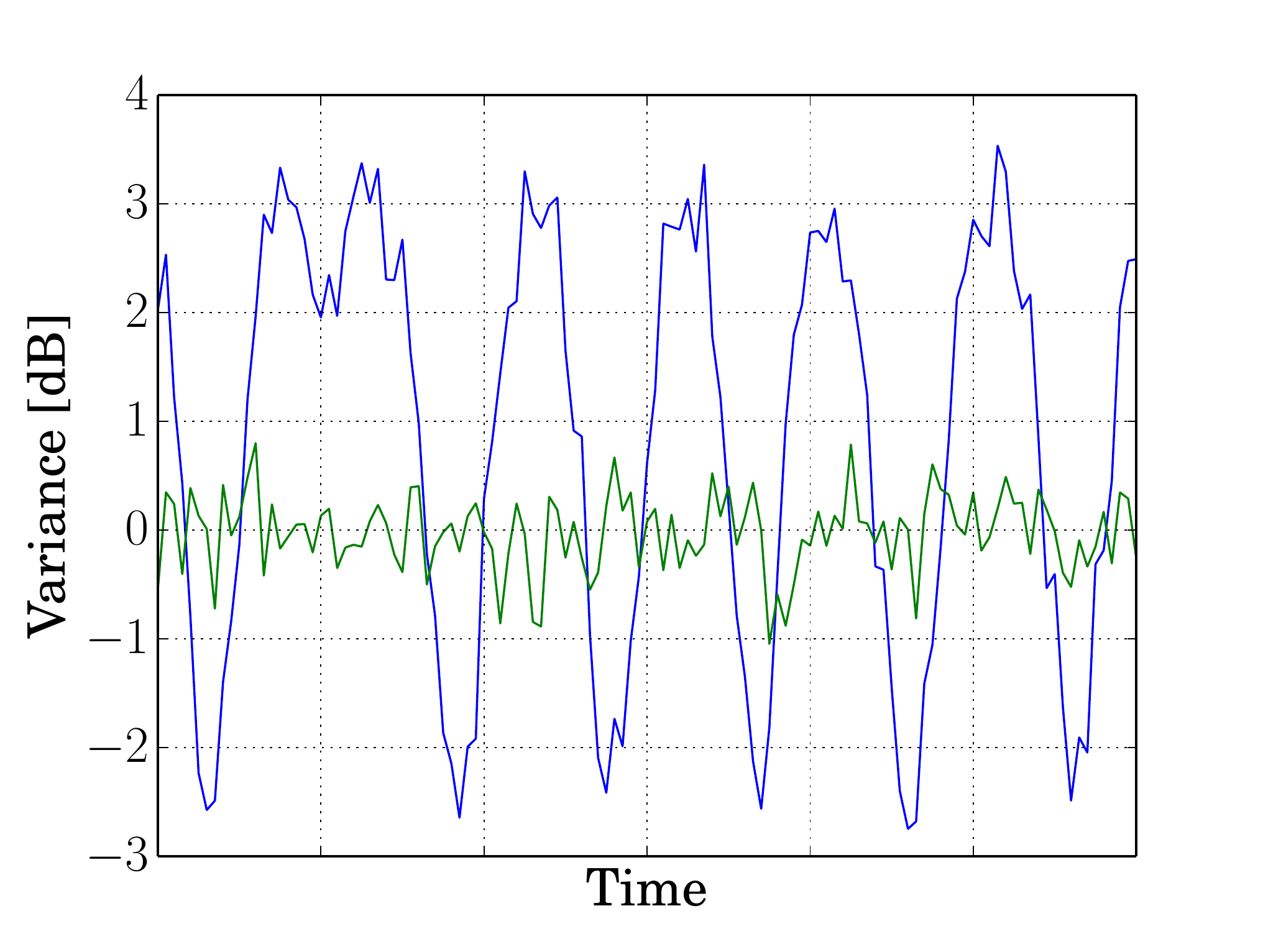}
\caption{Variance of the squeezed state used for the quantum gate, with a scanned local oscillator. The green trace represents shotnoise.}
\label{fig:SqueezingSupp}
\end{figure}

After entering the Faraday rotator from the other direction, the beam was now reflected, rather than transmitted, on the polarizing beam splitter. It then went through a half-wave plate and polarizing beam splitter combination with the same setting as the initial one, to simulate identical channels to and from the server. It was then interfered with the decryption beam on another beam sampler, to minimize loss to the signal. The two beams were locked to destructive interference by another sideband lock, using the same 36.7 MHz sideband, with a photodiode monitoring the interference fringes in the secondary output port of the beam sampler. The offset on this error signal allowed for optimizing the relative phase between the beams, and this, in addition to the adjustable gain settings for the correlated outputs, made it possible to cancel the encryption modulation very accurately. Following the decryption operation the signal was interfered with the local oscillator for scanned homodyne detection for density matrix reconstruction. Data acquisition was the same as for the displacement gate experiment. To compute the fidelity the definition in Equation~\eqref{eq:Fidelity} was used. To reconstruct the encrypted states, which are highly thermal with a slowly decaying photon distribution, a maximum likelihood algorithm which requires a truncation of the Fock space of the density matrix is not feasible. Instead the Wigner function was obtained directly from the data with the help of the inverse Radon transform and the filtered back-projection algorithm~\cite{Lvovsky2009}. For states of smaller magnitude where one can justify the truncation of the density matrix, the maximum likelihood algorithm is far superior~\cite{Lvovsky2009}. A plot of the scanned squeezing, obtained by homodyne detection, is shown in Fig.~\ref{fig:SqueezingSupp}.

We here list the factors contributing to optical loss in the system. Firstly, the mode matching of the input beam into the squeezer was 97 \%. This was partly because of a very high sensitivity to the focus of the incoming beam, but also because, as mentioned, the H\"ansch-Couillaud lock required some deviation from the ideal polarization. This intentional misalignment directly translated into loss of squeezing. Secondly, the reflected squeezed beam encountered a beam sampler, which directly produced 2 \% of transmission loss, but was necessary for the locking scheme, as the pump phase error signal was not sufficiently strong in the transmission of the squeezing cavity. Further, the H\"ansch-Couillaud lock clearly only works in reflection. The Faraday rotator induced 3 \% loss, and the beam sampler for interfering the signal with the decryption beam produced another 2 \% of loss. Lastly, the visibility of the homodyne detection was above 98 \% and the photodiodes in the homodyne detector both had a quantum efficiency of 99 \%.

\end{document}